\documentclass{aa}  

\usepackage{graphicx}

\usepackage{txfonts}

\usepackage{color}

\def\Rsun{{R}_{\sun}}
\def\Lsun{{L}_{\sun}}
\def\Msun{{M}_{\sun}}
\def\Tmean{\overline{T}_\mathrm{eff}}

\def\gmeanrbeta{\overline{g}_{\beta}}
\def\gmeanrelr{\overline{g}_\omega}
\def\diameq{\diameter_\mathrm{eq}}
\def\diamp{\diameter_\mathrm{p}}
\def\PArot{PA_\mathrm{rot}}
\def\vsini{V_\mathrm{eq} \sin i} 
\def\kms{\mathrm{km\,s}^{-1}}
\def\chir{\chi_\mathrm{r}}

\def\micron{\mu\mathrm{m}}
\def\Bproj{B_\mathrm{proj}}
\newcommand{\teff}{T_\mathrm{eff}}

\begin{document}

   \title{The evolved fast rotator Sargas}

   \subtitle{Stellar parameters and evolutionary status from VLTI/PIONIER and VLT/UVES\thanks{Based on observations performed at ESO, Chile under program IDs 097.D-0230(ABC) and 266.D-5655(A).}}



\authorrunning{A. Domiciano de Souza et al.}
\titlerunning{Sargas}

   \author{A. Domiciano de Souza\inst{\ref{loca}}\fnmsep\thanks{\email{Armando.Domiciano@oca.eu}}
          \and
          K. Bouchaud\inst{\ref{loca}, \ref{lesia}}
          \and
          M. Rieutord\inst{\ref{irap1}, \ref{irap2}}
          \and
          F. Espinosa Lara\inst{\ref{alcala}}
          \and
          B. Putigny\inst{\ref{irap1}, \ref{irap2}}
          }

   \institute{{Universit{\'e} C{\^o}te d'Azur, Observatoire de la C\^ote
    d'Azur, CNRS, UMR7293 Lagrange, 28 Av. Valrose, 06108 Nice Cedex 2, France\label{loca}}
          \and
              {LESIA, Observatoire de Paris, Universit{\'e} PSL, CNRS, Sorbonne Universit{\'e}, Univ. Paris Diderot, Sorbonne Paris Cité, 5 place Jules Janssen, 92195 Meudon, France \label{lesia}}    
         \and
             {Universit{\'e} de Toulouse, UPS-OMP, IRAP, Toulouse, France \label{irap1}}
          \and
             {CNRS, IRAP, 14, avenue Edouard Belin, 31400 Toulouse, France \label{irap2}}
         \and
             {University of Alcal{\'a}, 28871, Alcal{\'a} de Henares, Spain \label{alcala}}
             }

   \date{Received ...; accepted ...}

 
  \abstract
   {Gravity darkening (GD) and flattening are important consequences of stellar rotation. The precise characterization of these effects across the H-R diagram is crucial to a deeper understanding of stellar structure and evolution.}
   {We seek to characterize such important effects on \object{Sargas} (\object{$\theta$ Scorpii}), an  evolved, fast-rotating, intermediate-mass ($\sim5\,M_\sun$) star, located in a region of the H-R diagram where they have never been directly measured as far as we know.}
   {We use our numerical model CHARRON to analyze interferometric (VLTI/PIONIER) and spectroscopic (VLT/UVES) observations through a MCMC model-fitting procedure. The visibilities and closure phases from the PIONIER data are particularly sensitive to rotational flattening and GD. Adopting the Roche approximation, we investigate two GD models: (1) the $\beta$-model ($T_\mathrm{eff} \propto g_\mathrm{eff}^\beta$), which includes the classical von Zeipel's GD law, and (2) the $\omega$-model, where the flux is assumed to be anti-parallel to $g_\mathrm{eff}$. }
   {Using this approach we measure several physical parameters of Sargas, namely, equatorial radius, mass, equatorial rotation velocity, mean $T_\mathrm{eff}$, inclination and position angle of the rotation axis, and $\beta$. In particular, we show that the measured $\beta$ leads to a surface flux distribution equivalent to the one given by the $\omega$-model. Thanks to our results, we also show that Sargas is most probably located in a rare and interesting region of the H-R diagram: within the Hertzsprung gap and over the hot edge of the instability strip (equatorial regions inside it and polar regions outside it because of GD). 
}
   {These results show once more the power of optical/infrared long-baseline interferometry, combined with high-resolution spectroscopy, to directly measure fast-rotation effects and stellar parameters, in particular GD. As was the case for a few fast rotators previously studied by interferometry, the $\omega$-model provides a physically more profound description of Sargas' GD, without the need of a $\beta$ exponent. It will also be interesting to further investigate the implications of the singular location of such a fast rotator as Sargas in the H-R diagram.  }

  
\maketitle

\section{Introduction}

\object{Sargas} is the Sumerian-origin name\footnote{Approved by the IAU-Working Group on Star Names (WGSN).} of the evolved, intermediate-mass star \object{$\theta$ Scorpii} (\object{HIP86228}, \object{HD159532}, \object{HR6553}), the third apparently brightest star (mag $V=1.86$) and one of the most southern stars in the Scorpius constellation. 

Sargas is classified as an F1III giant by \citet[][adopted by the SIMBAD database]{Gray1989_v69p301-321}, while other authors classify Sargas as an F0-1II bright giant  \citep[e.g.,][]{Hohle2010_v331p349, Snow1994_v95p163-299, Hoffleit1991BrightStarCatalogue, Samedov1988_v28p335}. Recently, \citet{Ayres2018_v854p95} has placed Sargas between F1III and a low-luminosity supergiant of type F0Ib. 

\citet{Eggleton2008_v389p869-879} detected a faint companion (visible magnitude 5.36) at a relatively large angular separation of $6.47\arcsec$. Early on, \citet{See1896_v142p43} reported the discovery of a companion at a similar separation, (mean values of $6.24\arcsec$ at a position angle of $321.5\degr$), but much fainter (approx. thirteenth magnitude). \citet{Ayres2018_v854p95} discusses more recent results concerning the existence of this putative companion.

Despite it being an evolved star, well beyond the main sequence, Sargas is rapidly rotating, with a projected equatorial rotation velocity $\vsini \sim 105-125\,\kms$ \citep[e.g.,][]{Gebocki2005_v560p571, Ochsenbein1987_v32p83}. Such a high rotation velocity in an evolved intermediate-mass star is rather atypical, and therefore Sargas provides us with a rare opportunity to further our knowledge on the effects of rotation on post-main sequence stars. In particular, because of its fast rotation rate, Sargas is expected to present (1) geometrical flattening and (2) gravity darkening (GD). 

These two effects can be directly constrained by optical/infrared(IR) long-baseline interferometry (OLBI). Indeed, since the beginning of this century, modern stellar interferometers have measured these two consequences of rotation in different stars across the H-R diagram: Altair \citep[$\alpha$ Aql, A7IV-V;][]{van-Belle2001_v559p1155-1164, Domiciano-de-Souza2005_v442p567-578, Peterson2006_v636p1087-1097, Monnier2007_v317p342-345}, Achernar \citep[$\alpha$ Eri, B3-6Vpe;][]{Domiciano-de-Souza2003_v407pL47-L50, Kervella2006_v453p1059-1066, Domiciano-de-Souza2014_v569pA10}, Regulus \citep[$\alpha$ Leo, B8IVn;][]{McAlister2005_v628p439-452, Che2011_v732p68-80}, Vega \citep[$\alpha$ Lyr, A0V;][]{Aufdenberg2006_v645p664-675, Peterson2006_v440p896-899, Monnier2012_v761pL3}, Alderamin \citep[$\alpha$ Cep, A7IV;][]{van-Belle2006_v637p494-505, Zhao2009_v701p209-224}, Rasalhague \citep[$\alpha$ Oph, A5III;][]{Zhao2009_v701p209-224}, and Caph \citep[$\beta$ Cas, F2IV;][]{Che2011_v732p68-80}.

The estimated average angular diameter of Sargas is large enough \citep[$\sim2.5-3.3$ mas; ][]{van-Belle2012_v20p51-99, Chelli2016_v589pA112} to be resolved by the presently operating stellar interferometers, in particular by the ESO-VLTI, which is also well located to observe this southern star. In the following we describe the results of our interferometry-based study of Sargas, adding it to the ongoing list of fast-rotators measured by OLBI. Because the spectral type of Sargas is completely different from these previously observed rapidly rotating stars, the present OLBI study provides a new insight into fast rotation effects in an unexplored region of the H-R diagram.

The interferometric and spectroscopic ESO-VLT(I) data used in this work are described in Sect.~\ref{obsdatared}. The adopted stellar-rotation model is presented in Sect.~\ref{model}. The physical and geometrical parameters of Sargas estimated in a model-fitting procedure are presented in Sect.~\ref{results}. A discussion of these results and the conclusions of this work are given in Sects.~\ref{discussion} and \ref{conclusion}, respectively.

   
\section{Observations and data reduction \label{obsdatared}}

\subsection{VLTI/PIONIER \label{pionier}}

On 11 nights between April and September, 2016, photons from Sargas (and some targets for calibration) were simultaneously collected using four 1.8-m Auxiliary Telescopes (AT) of the ESO-VLTI \citep{Haguenauer2010_v7734p} and combined with the PIONIER instrument \citep{Le-Bouquin2011_v535pA67} giving rise to six interference fringe patterns per observation. 

These observations, summarized in Table~\ref{ta:log_pionier}, were reduced and calibrated using the standard PIONIER pipeline \textit{pndrs} \citep{Le-Bouquin2011_v535pA67}. After this procedure, each set of six raw fringes from Sargas was converted into a set of six calibrated squared visibilities $V^2$ and four calibrated closure phases $CP$ dispersed over six spectral channels (spectral bin width $\Delta\lambda=0.048\,\micron$) within the H band: wavelengths ($\lambda$) roughly centered at 1.52, 1.57, 1.62, 1.67, 1.72, and 1.76 $\micron$ (small variations of $\sim\pm0.01\,\micron$ exist between different observations). 

These final, calibrated data of Sargas are provided by the \texttt{OiDB}/JMMC\footnote{Jean-Marie Mariotti Center} service as 30 standard OIFITS files. After removing three individual $CP$ data points flagged as bad in these OIFITS files, our final dataset amounts to 1080 $V^2$ and 702 $CP$ points, that is, 1782 individual interferometric data points and corresponding uncertainties.

\begin{figure*}[ht]
\centering
\includegraphics[width=0.95\hsize]{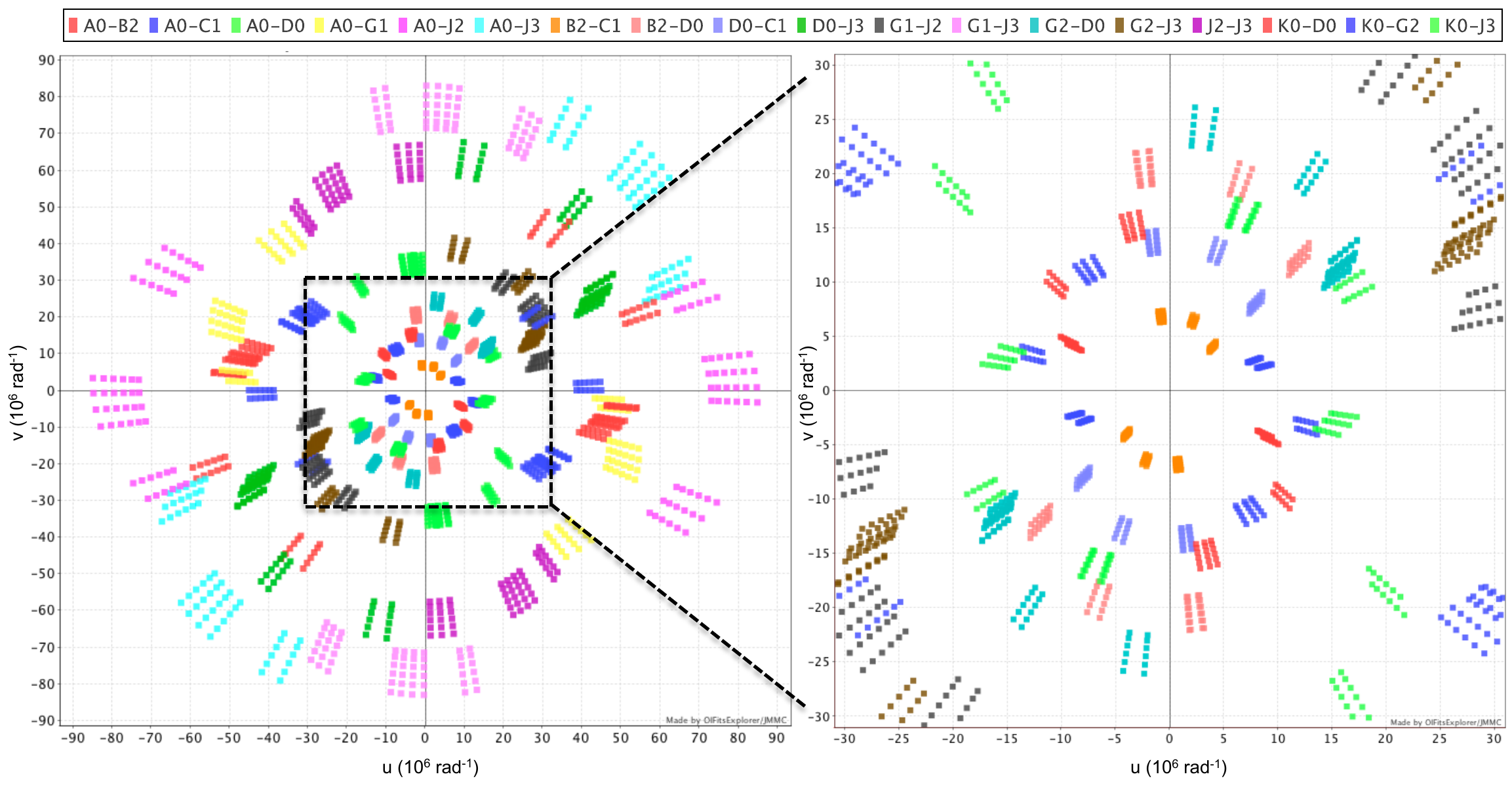}
\caption{{\it uv} coverage (units of spatial frequency) of VLTI/PIONIER observations of Sargas. The VLTI baselines used are identified with different colors (the corresponding AT stations are indicated in the upper legend). The six points per baseline correspond to the six spectral channels of the PIONIER H band observations as described in Sect.~\ref{pionier}. Projected baselines $\Bproj$ range from $\sim9$~ to $\sim132$~m. A zoom of the central region is shown in the right panel, corresponding to a maximum $\Bproj$ of $\sim50$~m. Image adapted from the \texttt{OIFits Explorer} service from JMMC.}
\label{fig:uv}
\end{figure*}

\begin{figure*}[!ht]  
\centering
 \includegraphics[width=0.46\hsize]{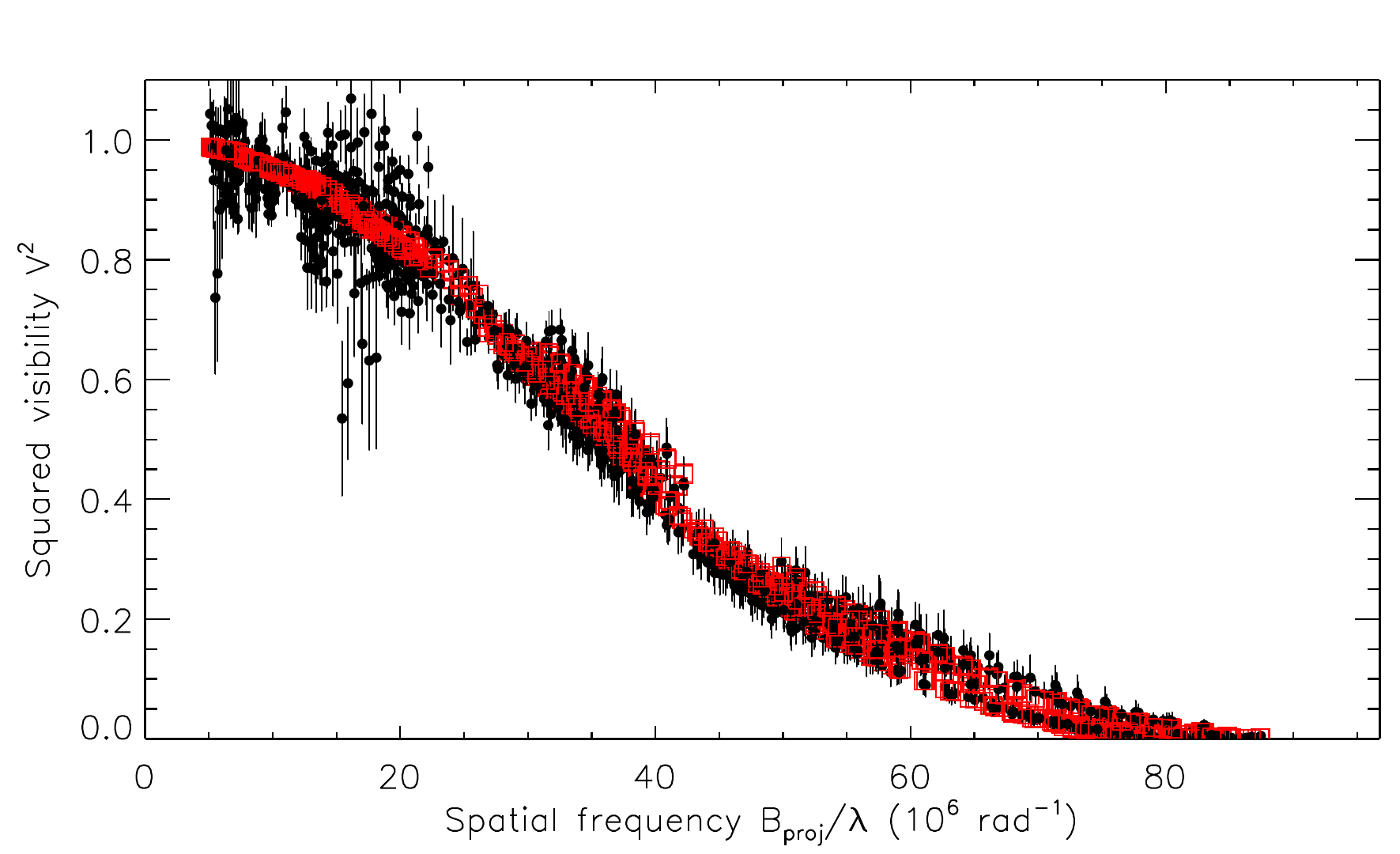}
\hspace*{0.3cm}
  \includegraphics[width=0.46\hsize]{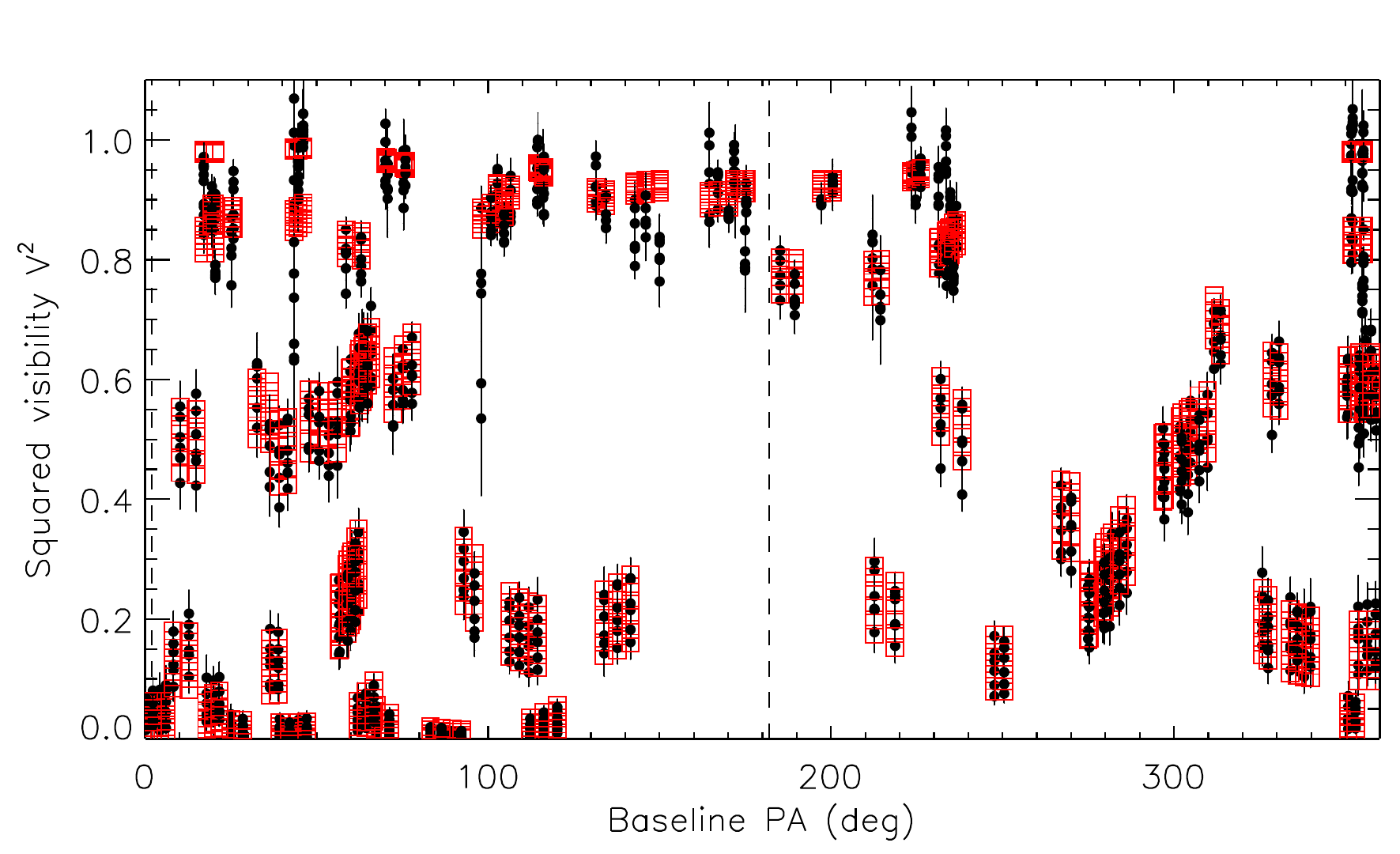}
  \includegraphics[width=0.46\hsize,]{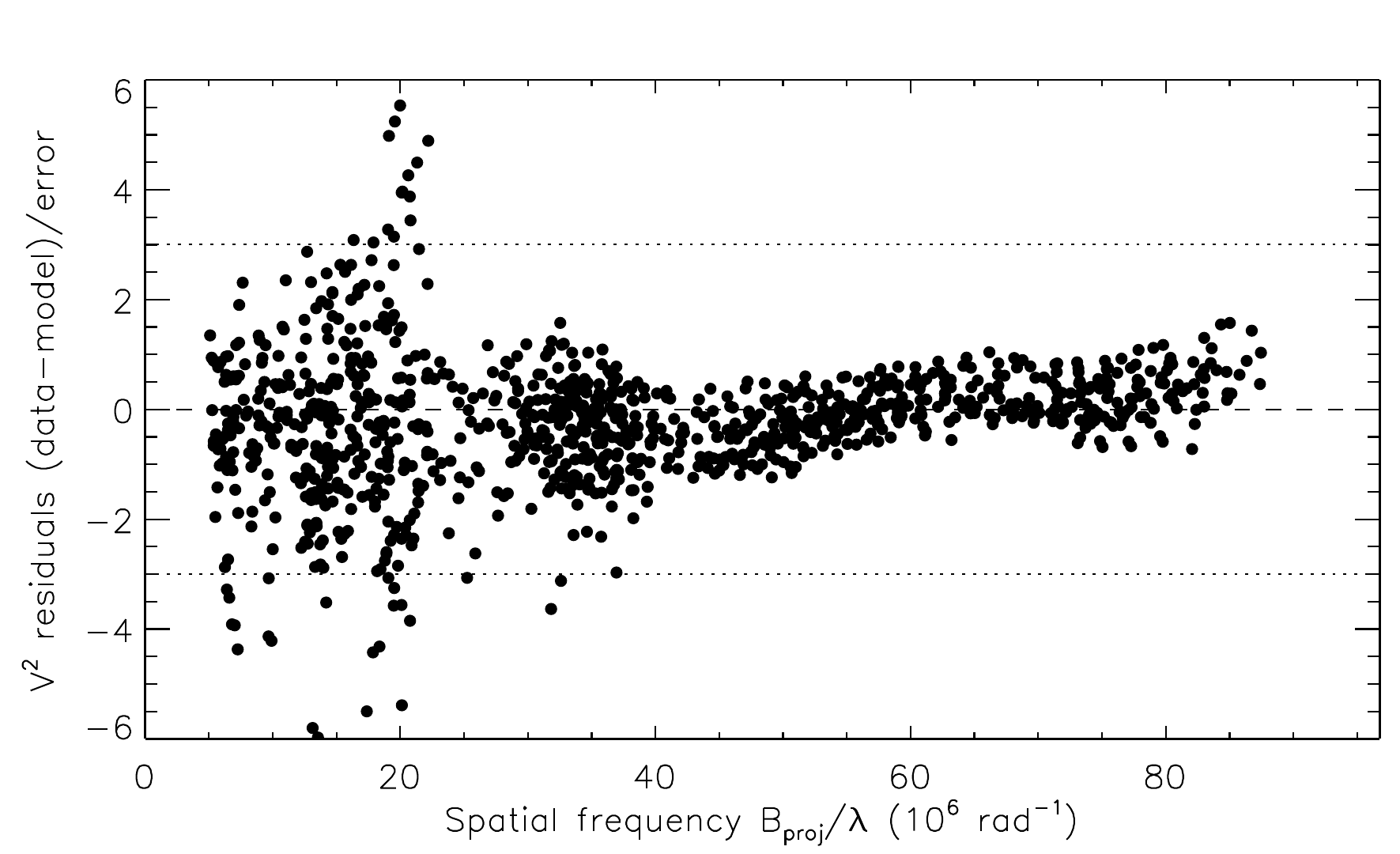}
\hspace*{0.3cm}
  \includegraphics[width=0.46\hsize]{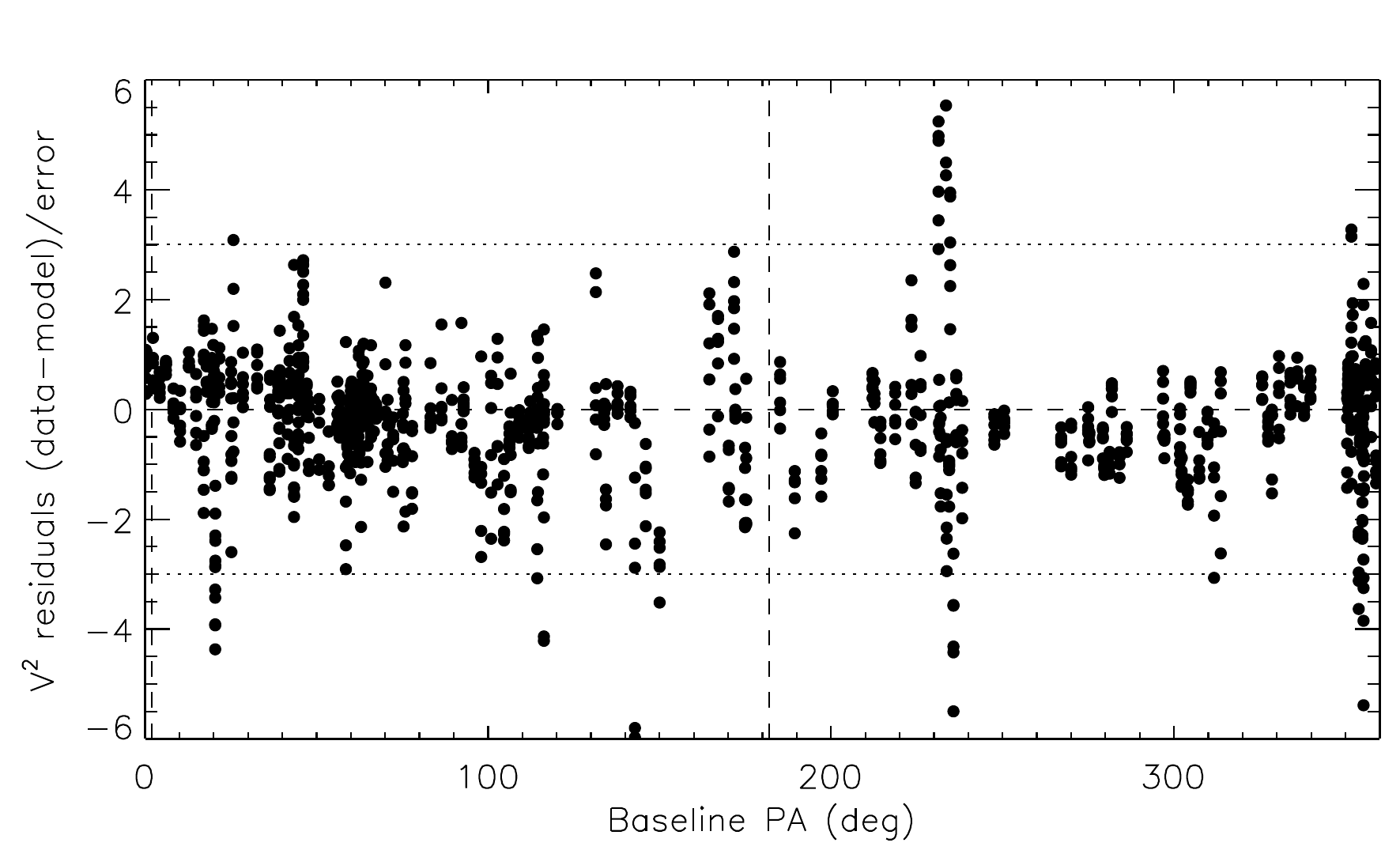}
   \includegraphics[width=0.46\hsize]{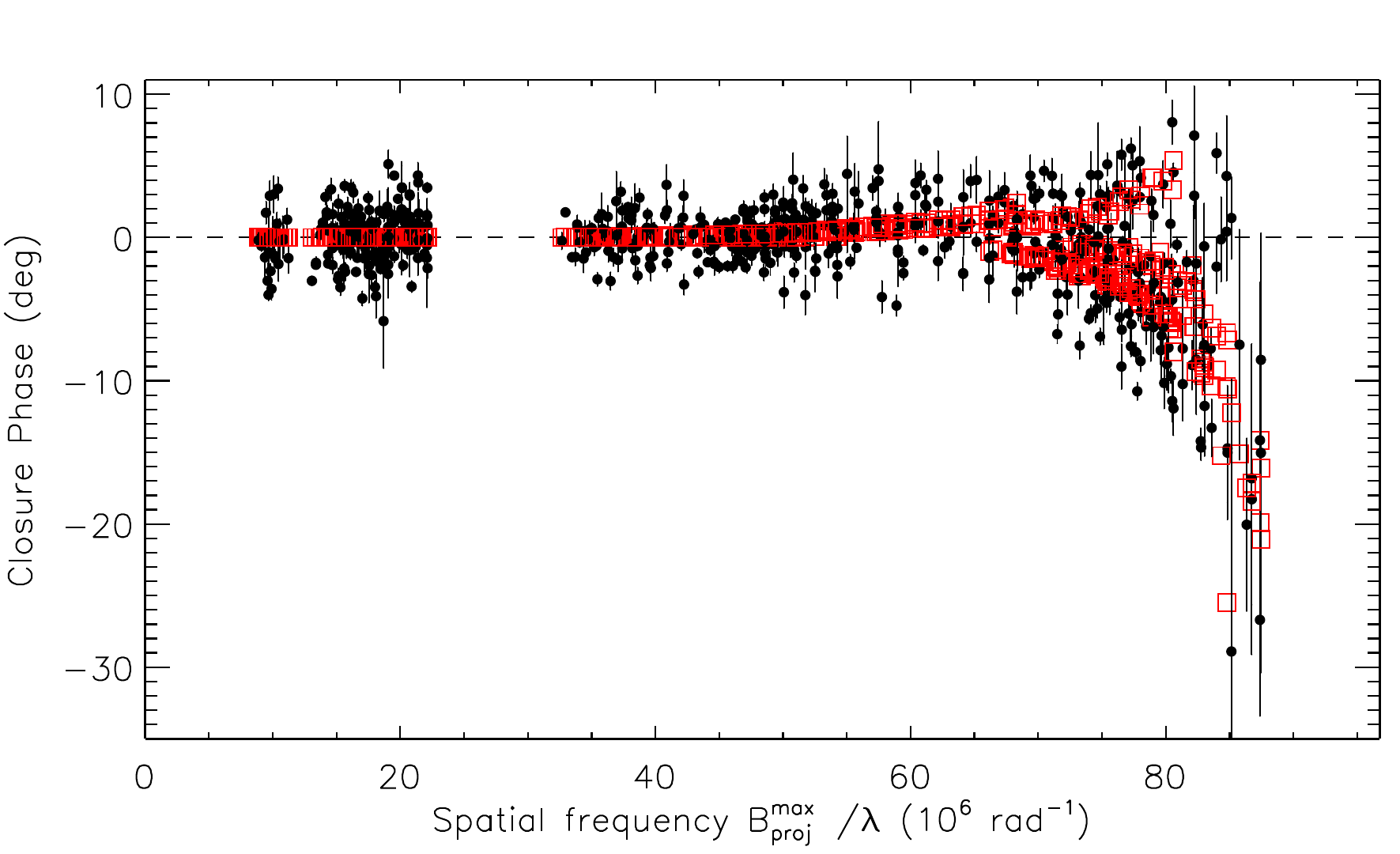}  
\hspace*{1.0cm}
   \includegraphics[width=0.46\hsize]{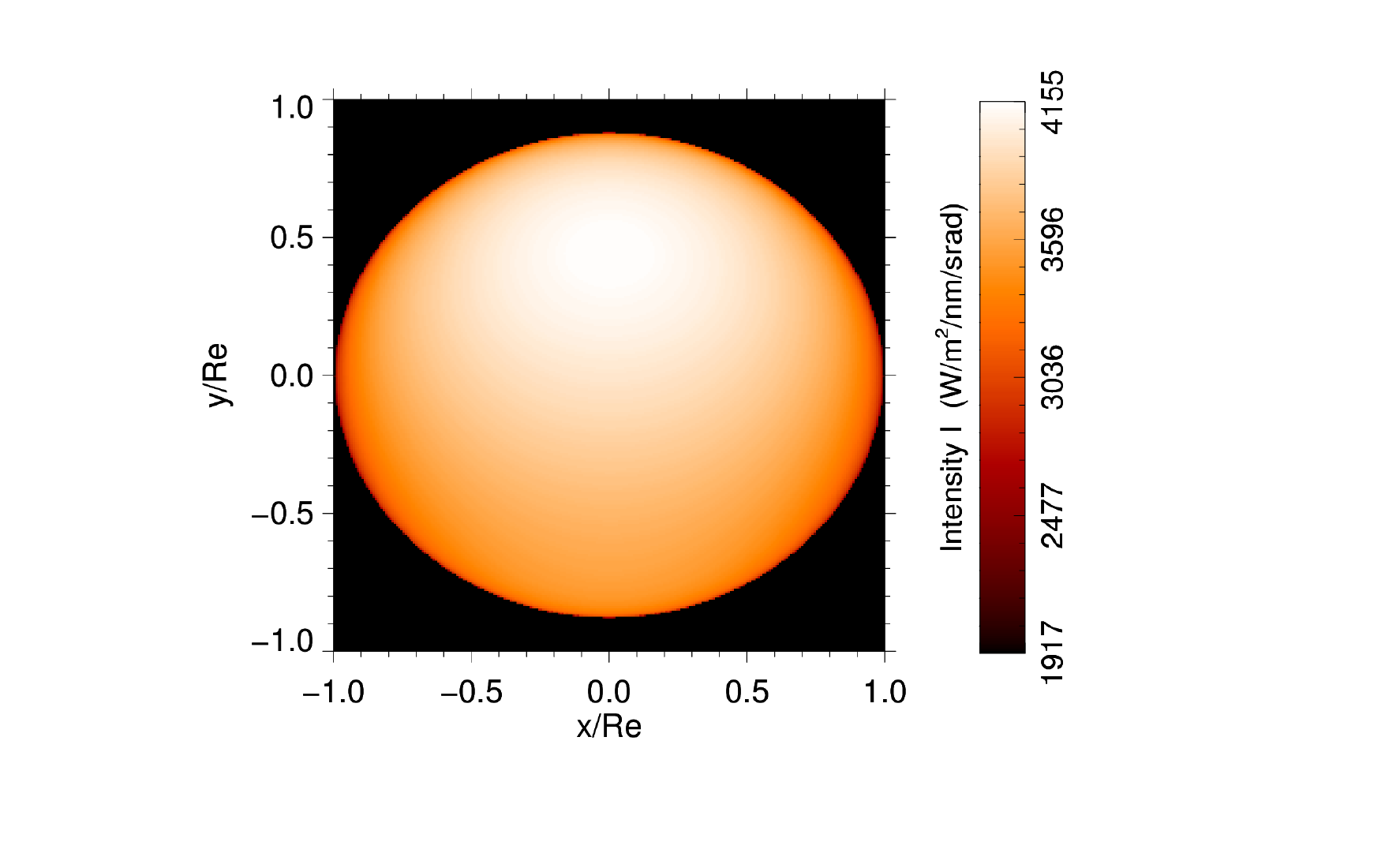}  
   \includegraphics[width=0.46\hsize]{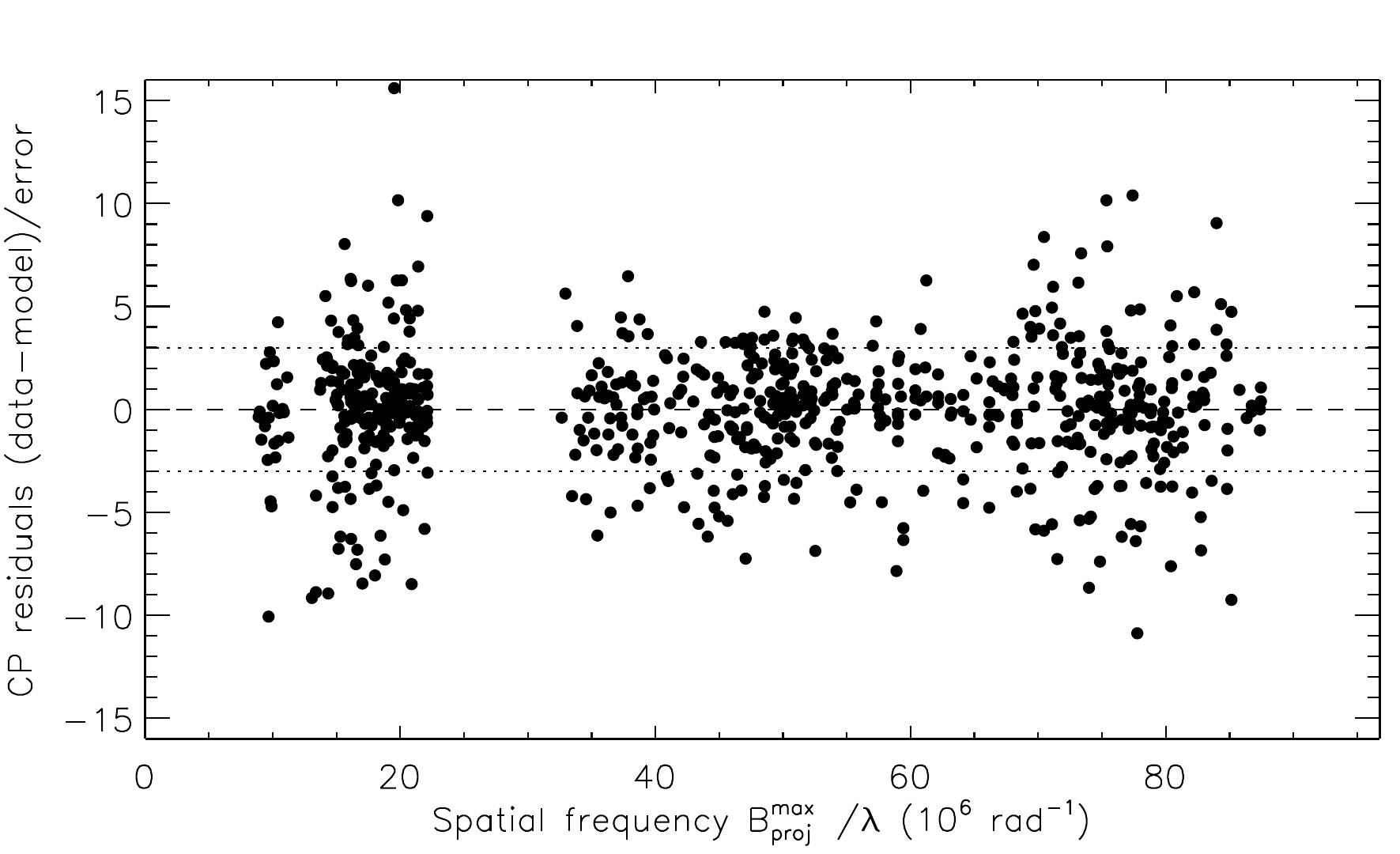}  
\hspace*{1.0cm}
   \includegraphics[width=0.46\hsize]{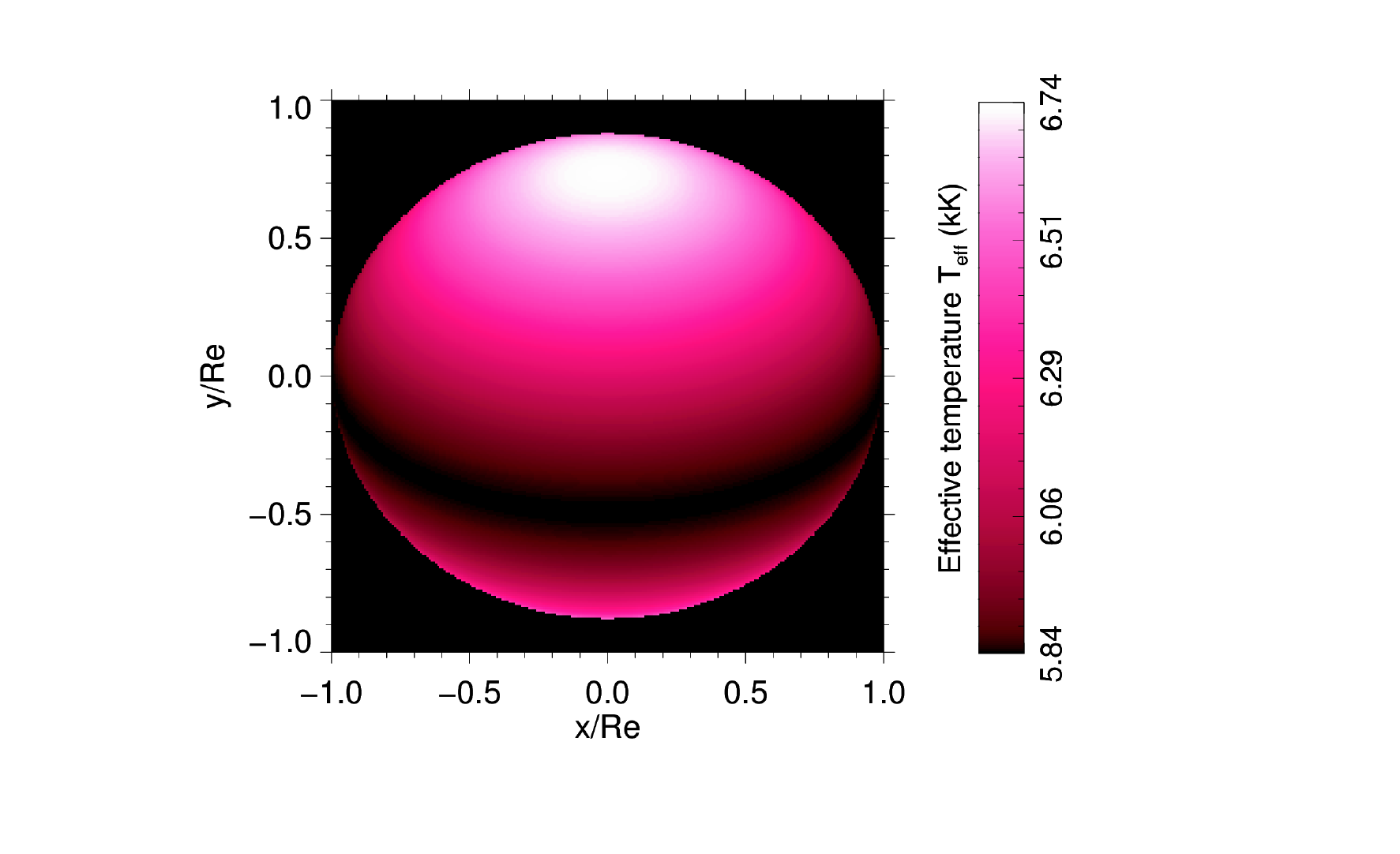}    
\caption{Comparison between interferometric observables from PIONIER (black dots with error bars) and from the $\beta$-best-fit model (red squares) presented in Table~\ref{ta:emcee_parameters}: squared visibilities $V^2$ as a function of both spatial frequencies and baseline position angles (PA), and closure phases as a function of spatial frequencies defined at the longest baseline of the corresponding triangle. The error-normalized residuals of these comparisons are also shown (black dots); horizontal lines indicate the 0 (dashed line) and $\pm3\sigma$ (dotted line) reference values. The vertical dashed lines in the plots with PA abscissas indicate the PAs of the sky-projected rotation axis for the visible ($\PArot$) and hidden ($\PArot-180\degr$) stellar poles. The two images at the bottom right show one specific intensity map, computed at the 1.62~$\micron$ spectral channel of the PIONIER data, and the $T_\mathrm{eff}$ map corresponding to this $\beta$-best-fit model. These images are not rotated by $\PArot$, i.e., they do not represent the star in sky coordinates. }
\label{fig:interf_data_charron_fit_Rbeta}
\end{figure*}

Figure~\ref{fig:uv} shows that the {\it uv}-plane of the PIONIER observations is well sampled, both in terms of position angles and projected baselines $\Bproj$, which range from $\sim9$~ to $\sim132$~m. Such good {\it uv} coverage and phase information are important assets in interferometric studies of fast rotators since these objects present non-centrosymmetric intensity distributions. The $V^2$ and $CP$ observables are shown in Figs.~\ref{fig:interf_data_charron_fit_Rbeta} and \ref{fig:interf_data_charron_fit_RELR} (Appendix~\ref{app:figs_relr}), which are further explained and discussed in the following sections, together with our analysis and modeling results.

Before finishing this section we would like to mention that the putative companion reported by \citet{See1896_v142p43} and by \citet{Eggleton2008_v389p869-879} should not induce any signature in the PIONIER AT observations since it is located well beyond ($\gtrapprox30$ times) the field-of-view of the instrument. In addition, considering the very faint (thirteenth) magnitude given by \citet{See1896_v142p43}, the putative companion would likely not impact the interferometric data even if it were nearby.

\begin{table}
\caption[]{\label{ta:log_pionier}Log of the VLTI/PIONIER observations of Sargas.}
\centering
\begin{tabular}{cccc}
\hline\hline
 Date             &   Nb.    &      AT        &    Calibration\\
(2016)            &  files   & configuration  &    stars    \\ 
\hline
Apr 5/6       &    2    &  A0-J2-G1-J3    &   HR6675            \\
Apr 6/7       &    2    &  A0-J2-G1-J3    &   HR6675            \\
May 22/23     &    7    &  A0-B2-C1-D0    &   HR6675, HR6783    \\
May 24/25     &    2    &  A0-B2-C1-D0    &   HR6675, HR6783    \\
May 27/28     &    2    &  A0-J2-G1-J3    &   HR6675, HR6783    \\
May 28/29     &    3    &  A0-J2-G1-J3    &   HR6675, HR6783    \\
May 30/31     &    4    &  D0-K0-G2-J3    &   HR6675, HR6783    \\
May 31/Jun 1  &    2    &  D0-K0-G2-J3    &   HR6675, HR6783    \\
Jul 1/2       &    2    &  D0-K0-G2-J3    &   HR6675, HR6783    \\
Aug 30/31     &    2    &  D0-K0-G2-J3    &   HR6675, HR6783    \\
Sep 1/2       &    2    &  D0-K0-G2-J3    &   HR6675, HR6783    \\
\hline
\end{tabular}
\end{table}

\subsection{VLT/UVES \label{uves}}


In addition to the interferometric data we also analyzed reduced spectroscopic data from VLT/UVES, made available by the ESO Science Archive Facility (SAF) in September 2013. We considered four similar spectra obtained on 10 September, 2002,  covering the visible wavelength range, from 472.6 to 683.5~nm, with a spectral resolution of $\sim74\,450$. A fifth spectrum taken on the same night was also available but was discarded as bad data, presenting a spurious sinusoidal signal. In our analysis, we choose a subset of the UVES spectra centered on the strong H$\alpha$ absorption line and covering the range between 643.2 and 669.2~nm, where other weaker lines are also present. 

The selected UVES spectra are not absolutely flux calibrated, and in order to use them in our analysis they were normalized and averaged as follows.
\begin{enumerate}
\item For the normalization process, the continuum regions were identified by comparing the observational data to synthetic spectra obtained from the AMBRE project \citep{de-Laverny2012_v544pA126, de-Laverny2013_v153p18-21}. Twelve wavelength values were selected that were deemed to correspond to continuum points around H$\alpha$. To determine the flux at these continuum wavelengths, a moving average was performed on the observational data so that the noise would not affect this continuum flux estimation. Then, for each of the four spectra, the continuum flux over the whole spectral range was obtained from a second-order polynomial fit over these twelve selected continuum points. Finally, this continuum flux was used to normalize both the flux and the associated uncertainties.
\item The average normalized flux $F_\mathrm{norm}$ (and uncertainty $\sigma_\mathrm{mean}$) was computed using the weighted average, since the four observed spectra do not have identical signal-to-noise ratios (S/N) (values ranging from $\sim360$ to $\sim470$).
\item The four normalized spectra do not perfectly overlap, even though the same 
normalization procedure was applied. This is because the continuum level is never identical among distinct observed spectra. The final total error on $F_\mathrm{norm}$ ($\sigma_{F_\mathrm{norm}}$) was therefore estimated by adding (quadratically) the standard deviation of the four spectra relative to $F_\mathrm{norm}$ to the uncertainty $\sigma_\mathrm{mean}$. These two uncertainties are of the same order of magnitude.
\end{enumerate}

Even though we limit the spectral range to a region close to H$\alpha$, the number of
wavelength points remains very large ($\simeq15\,000$) as a consequence of the high spectral resolution of the UVES data. Modeling a spectrum with such a huge number of wavelengths would be extremely time consuming and not suitable to our analysis (cf. Sects.~\ref{model} and \ref{results}).

To overcome this difficulty we considered a limited subset of 257 selected wavelengths, which ensured an acceptable computation time for the models, and at the same time preserved the high-spectral-resolution information in the UVES data. Indeed, the observed spectral lines are wide because of Sargas' high $\vsini$, which allowed us to determine this limited wavelength subset while keeping the physical information in the line profiles. This selected spectrum was also spectrally shifted based on the theoretical wavelength value of the H$\alpha$ line in order to be consistent with our analysis using modeled (not shifted) spectra. The 257 selected wavelengths homogeneously sample the whole considered spectral range (around H$\alpha$), with a denser sampling (by a factor $\simeq1.6$) in the regions of photospheric absorption lines.

\begin{figure*}[ht]
\centering
\includegraphics[width=0.85\hsize]{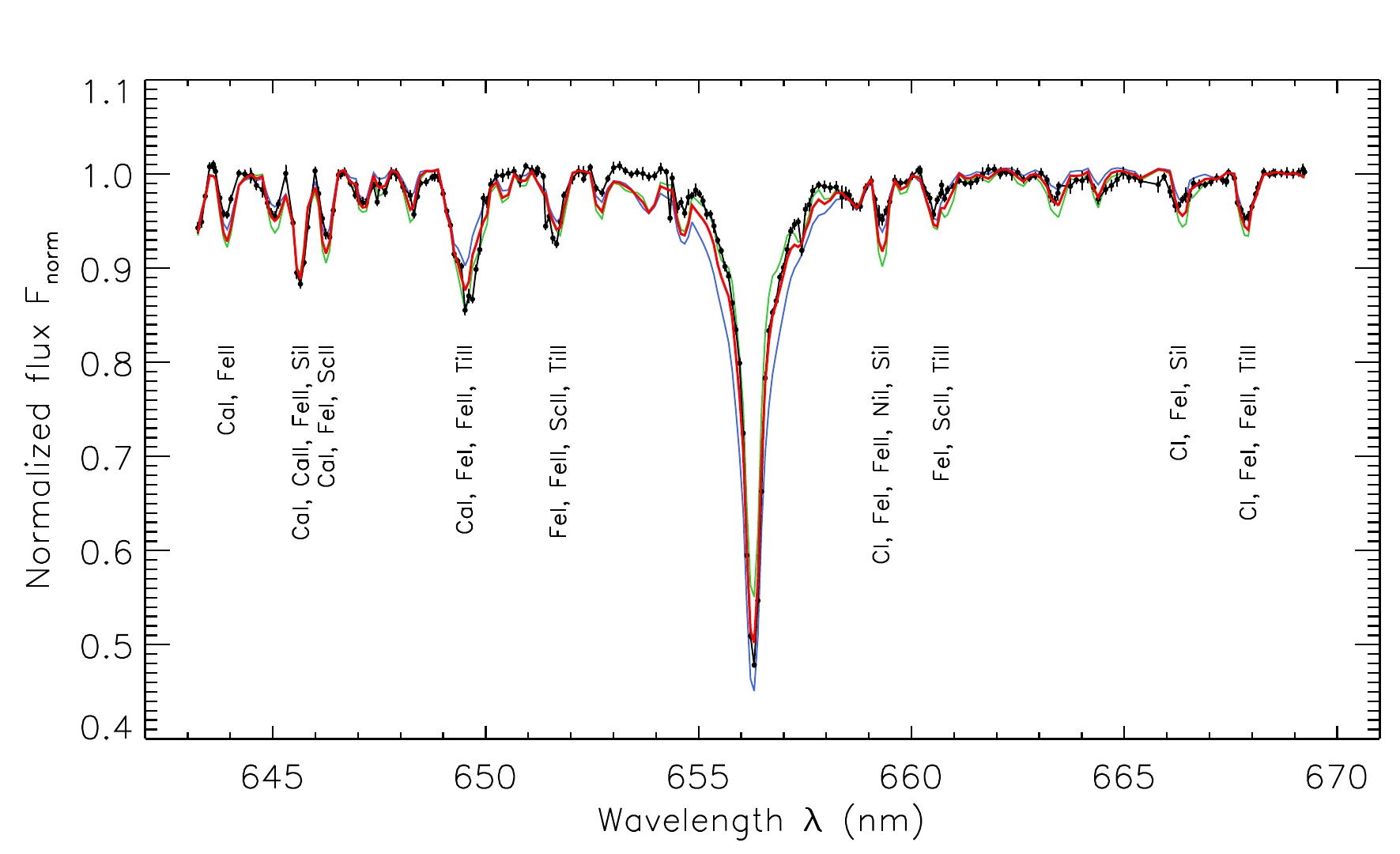}
\caption{Comparison between the normalized flux from UVES (black dots with error bars) and from the $\beta$-best-fit model (red curve) presented in Table~\ref{ta:emcee_parameters}. The spectra shown span $\sim 26$~nm centered on the H$\alpha$ line, where the main atoms and ions contributing to the strongest absorption lines have been identified. These observations correspond to 257 selected wavelengths, which 
homogeneously sample the whole relevant spectral range, as shown by the data points. The thin curves correspond to normalized model fluxes for the same $\beta$-best-fit model, but where we have fixed $T_\mathrm{eff}$ to $T_\mathrm{p}$ (blue) and $T_\mathrm{eq}$ (green) over the whole photosphere (model without GD, i.e., $\beta=0$). Clearly the complete best-fit model, with GD, better reproduces the observed spectral lines compared to the simpler model spectra computed with $\beta=0$.}
\label{fig:flux_data_charron_fit_Rbeta}
\end{figure*}

The final normalized and selected UVES spectrum around H$\alpha$ considered in this work is shown in Figs.~\ref{fig:flux_data_charron_fit_Rbeta} and \ref{fig:flux_data_charron_fit_RELR}, which are further explained and discussed in the following sections.

Before continuing we note that, after conducting our work using the reduced spectra released in 2013, a new release was delivered by ESO SAF in October 2017, which included some improvements in their processing method. Having already done our analysis with the 2013 release, we decided to compare the two releases to check whether the analysis had to be done again with the new release or not. We found that for almost the whole considered
spectral range around H$\alpha,$  the relative difference between the two spectra did not exceed 0.25\%. The difference reached 2.5\% at very few ($\sim3-4$) narrow wavelength regions, which also correspond to regions where the errors associated with the spectra are the highest. This led us to conduct our analysis with the spectra of the first release, since the small discrepancies between the two releases would hardly affect the results of this work.

%


%
%

\section{\label{model} Stellar-rotation model}

\subsection{\label{roche} Flattened Roche-star}

To interpret the foregoing observations of Sargas we adopt a stellar-rotation model similar to the one used by \citet{Domiciano-de-Souza2014_v569pA10}. The photospheric structure is given by the \textit{Roche model}: uniform (rigid) rotation with constant angular velocity $\Omega$ and mass $M$ concentrated at the center of the star. This is a very good approximation for such an intermediate mass, fast-rotating evolved star like Sargas. 

The rotationally flattened stellar photospheric surface follows the Roche equipotential so that the stellar radius $R$ as a function of the colatitude $\theta$ can be expressed as \citep[e.g.,][and references therein]{Kopal1987_v133p157-175, Domiciano-de-Souza2002_v393p345-357}

\begin{equation}
\label{eq:Rroche}
\begin{aligned}
R(\theta) &= R_\mathrm{eq}(1-\epsilon) \cfrac{\sin \left[\cfrac{1}{3}\arcsin
\left(  \cfrac{\Omega}{\Omega_\mathrm{c}}\sin \theta  \right) \right] }
{\cfrac{1}{3}\left( \cfrac{\Omega}{\Omega_\mathrm{c}}\sin \theta \right)} 
\end{aligned}
,\end{equation}
where the flattening parameter $\epsilon$ is given by
\begin{equation}
\label{eq:flattening}
\epsilon \equiv 1-\frac{R_\mathrm{p}}{R_\mathrm{eq}}=\frac{V_\mathrm{eq}^{2}R_\mathrm{p}}{2GM}
=\left(1+\frac{2GM}{V_\mathrm{eq}^{2}R_\mathrm{eq}}\right)^{-1} \ .
\end{equation}

In the foregoing equations, $G$ is the gravitational constant, $R_\mathrm{eq}$ and $R_\mathrm{p}$ are the equatorial and polar radii, $V_\mathrm{eq}(=\Omega R_\mathrm{eq})$ is the equatorial rotation velocity, and $\Omega_\mathrm{c}$ is the critical angular velocity of the Roche model, that is,
\begin{equation}
\Omega_\mathrm{c}=\frac{V_\mathrm{c}}{R_\mathrm{c}} = \sqrt{\frac{GM}{R_\mathrm{c}^3}} \ ,
\end{equation}
with $R_\mathrm{c}(=1.5R_\mathrm{p})$ being the critical equatorial radius. At critical rotation, $\epsilon$ reaches its maximum value of $1/3$. We also introduced the critical equatorial velocity $V_\mathrm{c}=\Omega_\mathrm{c}R_\mathrm{c}$. The ratio $\Omega/\Omega_\mathrm{c}$ in Eq.~\ref{eq:Rroche}, and the corresponding ratio $V_\mathrm{eq}/V_\mathrm{c}$ can be expressed in terms of $\epsilon$ as
\begin{equation}
\frac{\Omega}{\Omega_\mathrm{c}} = \frac{3}{2}(1-\epsilon)\sqrt{3\epsilon} \,\,\,\, \mathrm{and} \,\,\,\, 
\frac{V_\mathrm{eq}}{V_\mathrm{c}} = \sqrt{3\epsilon} \ .
\end{equation}
%


The Keplerian orbital angular and linear velocities ($\Omega_\mathrm{k}$ and $V_\mathrm{k}$) are also often used to scale $\Omega$ and $V_\mathrm{eq}$:
\begin{equation}
\label{eq:omega_epsilon}
\omega=\frac{\Omega}{\Omega_\mathrm{k}} = \frac{V_\mathrm{eq}}{V_\mathrm{k}} = \sqrt{\frac{2\epsilon}{1-\epsilon}}  \ ,
\end{equation}
where
\begin{equation}
V_\mathrm{k}=\Omega_\mathrm{k}R_\mathrm{eq} = \sqrt{\frac{GM}{R_\mathrm{eq}}} \ .
\end{equation}

From the gradient of the Roche potential we can derive the surface effective gravity, which, in spherical coordinates (unit vectors $\hat{r}, \hat{\theta}, \hat{\phi}$), is given by
\begin{equation}
\label{eq:vec_geff_spherical}
\vec{g}_\mathrm{eff}(\theta) =\left(-\frac{GM}{R^{2}(\theta)}+ 
\frac{V_\mathrm{eq}^{2}}{R_\mathrm{eq}^{2}} R(\theta) \sin^{2}\theta, \frac{V_\mathrm{eq}^{2}}{R_\mathrm{eq}^{2}} R(\theta)\sin\theta \cos\theta, 0\right) \ .
\end{equation}

The ratio between the equatorial $g_\mathrm{eq}$ and polar $g_\mathrm{p}$ surface effective gravities can be expressed as \citep[cf.][]{Domiciano-de-Souza2014_HDR}
\begin{equation}
\label{eq:gegp}
\frac{g_\mathrm{eq}}{g_\mathrm{p}} =  (1-\epsilon)(1-3\epsilon) \  .
\end{equation}

We note that the stellar flattened shape and effective gravity in the Roche model are totally defined by $R_\mathrm{eq}$, $M$, and $V_\mathrm{eq}$.



\subsection{\label{gd} Gravity darkening}

To analyze Sargas we consider two distinct GD prescriptions, which have been used in recent interferometry-based works on fast rotators. Both prescriptions assume that the stellar shape is given by the Roche model described in Sect.~\ref{roche}.

\subsubsection{$\beta$-model}

The first GD prescription is the $\beta$-model, which is a generalization of the classical von Zeipel law \citep{von-Zeipel1924_v84p665-683}. The local, photospheric radiative-flux is assumed to follow a power law of the local effective gravity $\vec{g}_\mathrm{eff}$, namely
\begin{equation}
\label{eq:von_zeipel_beta}
F(\theta) = \sigma  T_\mathrm{eff}^4(\theta)= C g_\mathrm{eff}^{4\beta}(\theta)  \,\, \Rightarrow \,\, 
T_\mathrm{eff}(\theta)=\left (\frac{C}{\sigma} \right )^{0.25} g_\mathrm{eff}^\beta(\theta) \ ,
\end{equation}
where $\sigma$ is the Stefan-Boltzmann constant. This equation allows us to compute the local effective temperature $T_\mathrm{eff}$ as a function of $g_\mathrm{eff}=\|\vec{g}_\mathrm{eff}\|$, but adds two new parameters, namely the GD exponent $\beta$ and the coefficient $C$. Instead of using $C$, we prefer to adopt the average effective temperature $\Tmean$, which is more directly related to observable quantities. $\Tmean$ and $C$ are related to the stellar luminosity $L$ by
\begin{equation}
\label{Tmean_L}
L=\sigma \int T_\mathrm{eff}^4(\theta)dS = \sigma \Tmean^4 S_\star 
= C\int g_\mathrm{eff}^{4\beta}(\theta)dS \equiv C  \gmeanrbeta^{4\beta} S_\star \, ,
\end{equation}
where $S_\star (=\int dS \equiv 4\pi\overline{R}^2)$ is the total stellar photospheric surface area, that is, the area of the Roche surface defined in Eq.~\ref{eq:Rroche}, and $\overline{R}$ is the radius of an equivalent spherical star of photospheric surface $S_\star$. In the above equation we have also introduced the $\beta$-model average gravity $\gmeanrbeta$, which is not a new quantity since it is obtained from Eqs.~\ref{eq:Rroche} and \ref{eq:vec_geff_spherical}. We can now express $T_\mathrm{eff}$ in the $\beta$-model as
\begin{equation}
\label{eq:teff_rbeta}
T_\mathrm{eff}(\theta)=\Tmean \left (\frac{g_\mathrm{eff}(\theta)}{\gmeanrbeta} \right )^\beta \ .
\end{equation}

The $\beta$-model is thus totally defined by five parameters, that is, three parameters describing the Roche model and two new parameters for the GD: $R_\mathrm{eq}$, $M$, $V_\mathrm{eq}$, $\Tmean$, and $\beta$.

\subsubsection{$\omega$-model}

As a second prescription for the GD, we adopt the $\omega$-model \citep{Espinosa-Lara2011_v533pA43}, which is also based on a Roche-star
model  and is detailed in \cite{Rieutord2016_v914p101}. Compared to the von Zeipel's law, which introduces the {\it ad hoc} parameter $\beta$, the $\omega$-model is based on the physical assumption that the flux and the effective gravity are anti-parallel in a radiative envelope. Namely, one assumes that
\begin{equation}\label{f_eq}
\vec{F}=-f(r,\theta)\vec{g}_\mathrm{eff}
,\end{equation}
where $f(r,\theta)$ is a universal function that only depends on the parameter $\omega$ (or, alternately, $\epsilon$) given by Eq.~\ref{eq:omega_epsilon}. The full ESTER  two-dimensional (2D) models of rapidly rotating early-type stars have shown that the assumption expressed by Eq.~\ref{f_eq} is actually a very good approximation even for very distorted stars, since the angle between the two vectors never exceeds half a degree \citep{Espinosa-Lara2011_v533pA43}. With this assumption, the $\omega$-model introduces no extra parameters and can give $\omega$ immediately.

From Eq.~\ref{f_eq} we deduce the following relation between the local flux (or $\teff$) and $g_\mathrm{eff}$.
\begin{equation}
F(\theta) = \sigma T_\mathrm{eff}^4(\theta) = f(r(\theta), \theta) g_\mathrm{eff}(\theta) \ .
\end{equation}
Here, $r(\theta)$ is the scaled (by $R_{eq}$) radial distance of the stellar surface and the function $f(r,\theta)$ is determined from the flux conservation equation $\nabla\cdot\vec{F} = 0$.
\citet{Espinosa-Lara2011_v533pA43} show how to solve this equation  to obtain $T_\mathrm{eff}$ in the $\omega$-model (their Eq.~31):
\begin{equation}
T_\mathrm{eff}(\theta)=\left (\frac{F}{\sigma} \right )^{0.25}=\left (\frac{L}{4\pi \sigma G M}\right )^{0.25} \sqrt{\frac{\tan \vartheta(r(\theta), \theta)}{\tan \theta} } \, g_\mathrm{eff}^{0.25} \ ,
\end{equation}
where $\vartheta$ is given by
\begin{equation}
\cos\vartheta+\ln\tan(\vartheta/2) =
\frac{1}{3}\omega^2r(\theta)^3\cos^3\theta+\cos(\theta)+\ln\tan(\theta/2) 
\end{equation}
at each colatitude  \citep[see][Eq.~18]{Rieutord2016_v914p101}.

As before, we use the average effective temperature $\Tmean$ to rewrite the equation above as
\begin{equation}
\label{eq:teff_relr}
T_\mathrm{eff}(\theta)=\Tmean \left( \frac{g_\mathrm{eff}(\theta)}{\gmeanrelr} \right)^{0.25} \sqrt{\frac{\tan \vartheta(r(\theta), \theta)}{\tan \theta} } \ .
\end{equation}

The parameter $\gmeanrelr$ is the average gravity, given by
\begin{equation}
\gmeanrelr = \frac{4\pi G M}{S_\star} = \frac{G M}{\overline{R}^2} \ ,
\end{equation}
which is distinct from $\gmeanrbeta$, but still derived from previously defined quantities.

The $\omega$-model requires only $\Tmean$ as an additional parameter to the three Roche-star parameters (no $\beta$ exponent is needed). It is therefore completely defined by $R_\mathrm{eq}$, $M$, $V_\mathrm{eq}$, and $\Tmean$.

Although the GD exponent $\beta$ is not required in the $\omega$-model, it is useful to define an equivalent exponent $\beta_\omega$ for a future comparison with the $\beta$-model. There are different ways to define $\beta_\omega$ and here we consider the expression from \citet[][Eq.~28]{Rieutord2016_v914p101}, rewritten in terms of $\epsilon$ thanks to Eq.~\ref{eq:omega_epsilon}:
\begin{equation}
 \label{eq:beta_ELR}
   \begin{aligned}
\beta_\omega &= \frac{\ln \left( T_\mathrm{eq}/T_\mathrm{p} \right)}{\ln \left( g_\mathrm{eq}/g_\mathrm{p} \right)}
    =\frac{1}{4} - \frac{1}{6} \frac{\ln \left(\frac{1-3\epsilon}{1-\epsilon} \right) +2\epsilon(1-\epsilon)^2} {\ln\left[(1-\epsilon)(1-3\epsilon)\right]} \\
    &\simeq  \frac{1}{4} - \frac{1}{3} \epsilon \ ,  
   \end{aligned}
,\end{equation}
where the approximate result in the last line above is a useful first-order approximation for $\beta_\omega$.

In the previous equation, the ratio $g_\mathrm{eq}/g_\mathrm{p}$ is given by Eq.~\ref{eq:gegp} and the equatorial to polar effective temperatures ratio is given by \citep[cf.][]{Espinosa-Lara2011_v533pA43, Domiciano-de-Souza2014_HDR}
\begin{equation}
 \label{eq:TeTp_ELR}
 \frac{T_\mathrm{eq}}{T_\mathrm{p}} = \sqrt{1-\epsilon}\left(\frac{1-3\epsilon}{1-\epsilon}\right)^{1/12}\exp{\left(-\frac{\epsilon(1-\epsilon)^2}{3}\right)} \ .
\end{equation}

\subsection{Images and observables with CHARRON}

Both the $\beta$- and the $\omega$-models were implemented in our IDL-based program  CHARRON\footnote{\textit{C}ode for \textit{H}igh \textit{A}ngular \textit{R}esolution of \textit{R}otating \textit{O}bjects in \textit{N}ature.}, briefly described here. A detailed description is given by \citet{Domiciano-de-Souza2002_v393p345-357, Domiciano-de-Souza2012_vp321-324} and \citet{Domiciano-de-Souza2014_HDR}.

In CHARRON, the 2D Roche photosphere is divided into nearly equal surface area elements ($\sim 50\,000$). From the input parameters and equations defining the $\beta$- and the $\omega$-models, CHARRON associates, with each area element, a set of relevant physical quantities, such as, effective temperature $T_\mathrm{eff}(\theta)$ and gravity $\vec{g}_\mathrm{eff}(\theta)$, rotation velocity $V(\theta)(=\Omega R(\theta))$, normal surface vector, among others.


From these physical quantities, CHARRON computes wavelength($\lambda$)-dependent intensity maps of the visible stellar surface (images) by associating a local specific intensity $I$ with each surface area element. Each local $I$ depends on several parameters: $g_\mathrm{eff}$, $T_\mathrm{eff}$, metallicity, microturbulent velocity, rotation velocity projected onto the observer's direction ($V_\mathrm{proj}$), and others. To compute specific intensity maps it is also necessary to define the observer's position relative to the modeled star. This requires the introduction of three "position" parameters, namely, the stellar distance $d$, the rotation-axis inclination angle $i$, and the position angle of the sky-projected rotation axis $\PArot$ (counted from north to east until the visible stellar pole).

In this work, the CHARRON images are thus obtained by associating, with each surface grid element $j$, a local specific intensity from a plane-parallel atmosphere model, so that
\begin{equation}
I_j=I_j(g_{\mathrm{eff},j}, T_{\mathrm{eff},j}, \lambda(V_{\mathrm{proj},j}), \mu_j) \ ,
\end{equation}
where $\lambda$ is Doppler-shifted to the local $V_{\mathrm{proj},j}$, and $\mu_j$ is the cosine between the normal to the surface grid element and the line-of-sight (limb-darkening is thus included). Both $V_{\mathrm{proj},j}$ and $\mu_j$ depend on the inclination $i$.

The $I_j$ are obtained by interpolating grids of synthetic spectra pre-calculated for different values of $g_\mathrm{eff}$, $T_\mathrm{eff}$, and $\lambda$. To model the high-resolution UVES spectra we used the synthetic fluxes grid from the AMBRE project \citep{de-Laverny2012_v544pA126, de-Laverny2013_v153p18-21} obtained from the MARCS model atmospheres for late-type stars \citep{Gustafsson2008_v486p951-970}. Since the AMBRE spectra do not cover the IR domain, we adopted the ATLAS9 grid of low-spectral-resolution fluxes from \citet{Castelli2004_astro-ph} in order to model the low-spectral-resolution PIONIER observations. We checked that the spectra from these two models agree well in their common domain (visible).

We adopted synthetic spectra computed with solar abundance for Sargas since it is a nearby star and is not known to be chemically peculiar. This choice is justified by the chemical composition measurements from \citet{Hinkel2014_v148p54}, who report abundances very close to solar. In particular, these authors give $\mathrm{[Fe/H]}=0$ and $\mathrm{[Ti/Fe]}=0$, which are elements contributing to most UVES spectral lines considered in this work. \citet{Samedov1988_v28p335} also reports a chemical composition close to but somewhat lower than the solar one. Moreover, we note that small deviations from the adopted chemical composition do not significantly impact the low-spectral-resolution PIONIER data, which are the main observational set in our analysis.

The adopted AMBRE and ATLAS9 spectral fluxes were computed with microturbulent velocities $\xi_\mathrm{t}$ of 1 and $2\,\kms$, respectively. A precise measure of this parameter on a fast rotator such as Sargas is not easy because line broadening is completely dominated by rotational Doppler broadening. We thus estimated $\xi_\mathrm{t}$ for Sargas from the empirical relation of \citet[][Eq.~A1]{Adibekyan2015_v450p1900-1915} for evolved K and G stars (similar but slightly colder than Sargas): $\xi_\mathrm{t}\simeq1.5-2.0\,\kms$. This estimation confirms the expected low value of $\xi_\mathrm{t}$ for Sargas and justifies the adopted value. In any case, this parameter does not play an important role in a fast rotator such as Sargas since (1) $V_\mathrm{eq} \sim50 \xi_\mathrm{t}$ (i.e., line broadening is dominated by rotation); and (2) $\xi_\mathrm{t}$ has negligible influence on the broad band interferometric observations, both because it hardly affects the continuum spectral distribution and  because it does not influence the overall shape (size) or alter the large-scale surface distribution of the star.

Since only the fluxes $F$ of the AMBRE and ATLAS9 spectral grids are available, and not the specific intensities, we converted $F$ into a corresponding grid of surface-normal specific intensities $I(\mu=1)$ using the four-parameter non-linear limb-darkening law from \citet{Claret2000_v363p1081-1190} (his Eq.~6). By interpolating $I(\mu=1)$ as explained in the previous paragraph and using Claret's limb-darkening law, CHARRON computes $I_j$ at each stellar surface area element to generate images (sky-projected specific intensity maps); examples are shown in Figs.~\ref{fig:interf_data_charron_fit_Rbeta} and \ref{fig:interf_data_charron_fit_RELR}, which are presented in the following sections. 

The CHARRON-simulated observed fluxes and interferometric observables (e.g., visibilities and closure phases) are directly obtained from the Fourier transform of these images \citep[e.g.,][]{Domiciano-de-Souza2002_v393p345-357}.



\section{\label{results} Sargas' parameters from MCMC model-fitting}

%
%
\begin{table*}[ht]
\caption[]{\label{ta:emcee_parameters} Physical and geometrical parameters of Sargas measured from the MCMC fit of the $\beta$- and $\omega$-models to the PIONIER ($V^2$ and $CP$) and UVES (flux) data using the CHARRON and emcee codes (further details in the text of Sect.~\ref{results}). The best-fit values correspond to the median of histograms representing the probability distributions of the fitted parameter, while the uncertainties correspond to the commonly used 15.9 ($-1\sigma$) and 84.1 ($+1\sigma$) percentiles (cf. Figs.~\ref{fig:emcee_histo_results_Rbeta} and \ref{fig:emcee_histo_results_RELR}). We also give derived parameters with additional useful information on Sargas, obtained from the fitted parameters and from the equations in Sect.~\ref{model}.}
\centering
\begin{tabular}{ccc}
\hline\hline
\textbf{Fitted parameters}                                              & \textbf{$\beta$-model}       & \textbf{$\omega$-model} \\
\hline
Equatorial radius: $R_\mathrm{eq}$ $(\Rsun)$                            & $30.30^{+0.09}_{-0.10}$       & $30.30^{+0.08}_{-0.09}$     \\
Stellar mass: $M$ $(\Msun)$                                             & $5.08^{+0.14}_{-0.14}$        & $5.09^{+0.13}_{-0.14}$ \\
Equatorial rotation velocity: $V_\mathrm{eq}$ ($\kms$)                  & $105.5^{+1.8}_{-1.8}$         & $104.0^{+0.9}_{-1.1}$   \\
Inclination angle of rotation-axis: $i$ ($\degr$)                               & $60.2^{+1.7}_{-1.6}$          & $61.8^{+0.8}_{-0.9}$     \\
Position angle of rotation-axis\tablefootmark{a}: $\PArot$ ($\degr$)    & $181.8^{+0.5} _{-0.3}$        & $182.1^{+0.5} _{-0.4}$  \\
Average effective temperature: $\Tmean$ (K)                             & $6207^{+9}_{-10}$             & $6215^{+7}_{-8}$ \\     
Gravity-darkening coefficient: $\beta$                                  & $0.192^{+0.008}_{-0.007}$     & $-$   \\
\hline
\textbf{Derived parameters}                                                           & \textbf{$\beta$-model}        & \textbf{$\omega$-model} \\
\hline
Equatorial angular diameter: $\diameq=2R_\mathrm{eq}/d$ (mas)                         &        3.09                           &    3.09  \\
Polar radius: $R_\mathrm{p}$ $(\Rsun)$                                                &       25.81                           &   25.92  \\
Polar angular diameter: $\diamp=2R_\mathrm{p}/d$ (mas)                                &        2.63                           &    2.64  \\
Equatorial-to-polar radii ratio: $R_\mathrm{eq}/R_\mathrm{p}$                         &       1.1741                          & 1.1689  \\
Flattening: $\epsilon = 1-R_\mathrm{p}/R_\mathrm{eq}$                                 &       0.14825                         & 0.14446 \\
Radius of equivalent spherical star with same surface: $\overline{R}~(\Rsun)$         &        28.62                          &    28.66  \\
Angular diameter of equivalent spherical: $\overline{\diameter}=2\overline{R}/d$ (mas)&        2.92                           &    2.93  \\
Projected rotation velocity: $\vsini$ $(\kms)$                                        &        91.5                           &    91.7 \\
Equatorial and polar $T_\mathrm{eff}$: $T_\mathrm{eq}$; $T_\mathrm{p}$ (K)            &        5836; 6740                     & 5841; 6770 \\
Equatorial and polar gravities: $\log g_\mathrm{eq}$; $\log g_\mathrm{p}$ (dex)       &        1.995; 2.320                   & 2.003; 2.317 \\
Luminosity: $L$ $(\Lsun)$; $\log L/\Lsun$                                             &        1091; 3.038                    & 1100; 3.041 \\
Rotation period and frequency:  $P_\mathrm{rot}$ (day);  $\Omega$ (rad/day)           &        14.54; 0.432                   & 14.74; 0.426 \\
Critical rotation rate (angular and linear): 
$\Omega/\Omega_\mathrm{c}$; $V_\mathrm{eq}/V_\mathrm{c}$                             &        0.852; 0.667                    & 0.845; 0.658 \\
Keplerian orbital rotation rate: $V_\mathrm{eq}/V_\mathrm{k}=
\Omega/\Omega_\mathrm{k}$                              &        0.590                           &      0.581 \\
Equivalent gravity darkening coefficient: $\beta_\omega$  &     $-$                            &    0.204  \\ 
%
%
\hline
\end{tabular}
\tablefoot{
\tablefoottext{a}{Position angle of the sky-projected rotation-axis, counted from north to east until the visible stellar pole.}
}
\end{table*}
\begin{figure*}[ht]
\centering
\includegraphics[width=\hsize]{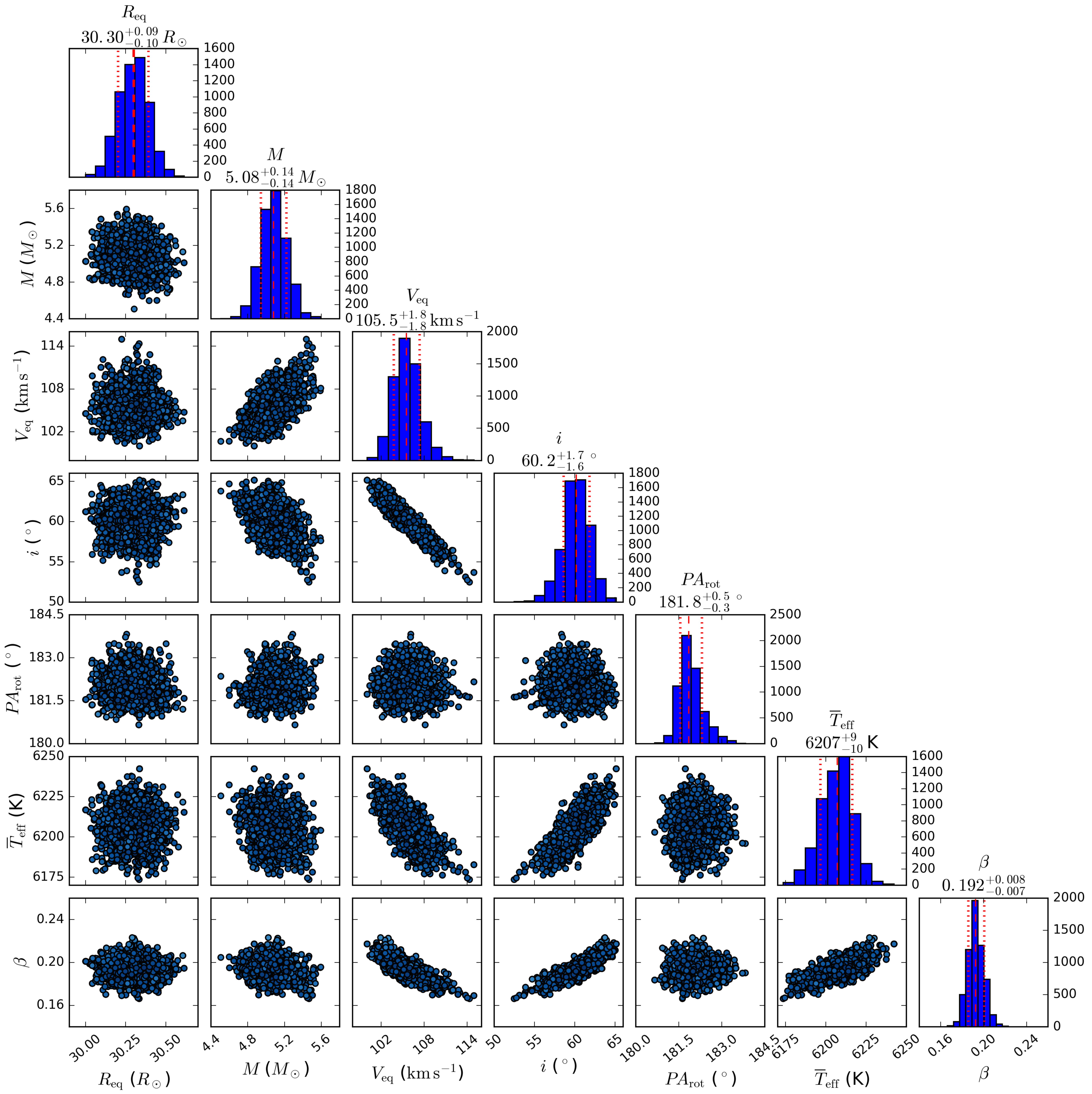}
\caption{Results obtained from the fit of the $\beta$-model to the interferometric and spectroscopic observations Sargas using the CHARRON and emcee codes (further details in the text). The figure shows one- (histograms) and two-dimensional projections of the posterior probability distributions obtained from the fitted parameters. The median (dashed lines) and uncertainties (dotted lines) obtained from the histograms for each fitted parameter are also indicated over the plots and correspond to the values given in Table~\ref{ta:emcee_parameters}. In the two-dimensional plots lower (higher) $\chi^2$ values correspond to darker (lighter) symbols; these 6000 points correspond to the converged final results of the MCMC model-fitting with emcee.}
\label{fig:emcee_histo_results_Rbeta}
\end{figure*}
\begin{figure*}[ht]
\centering
\includegraphics[width=\hsize]{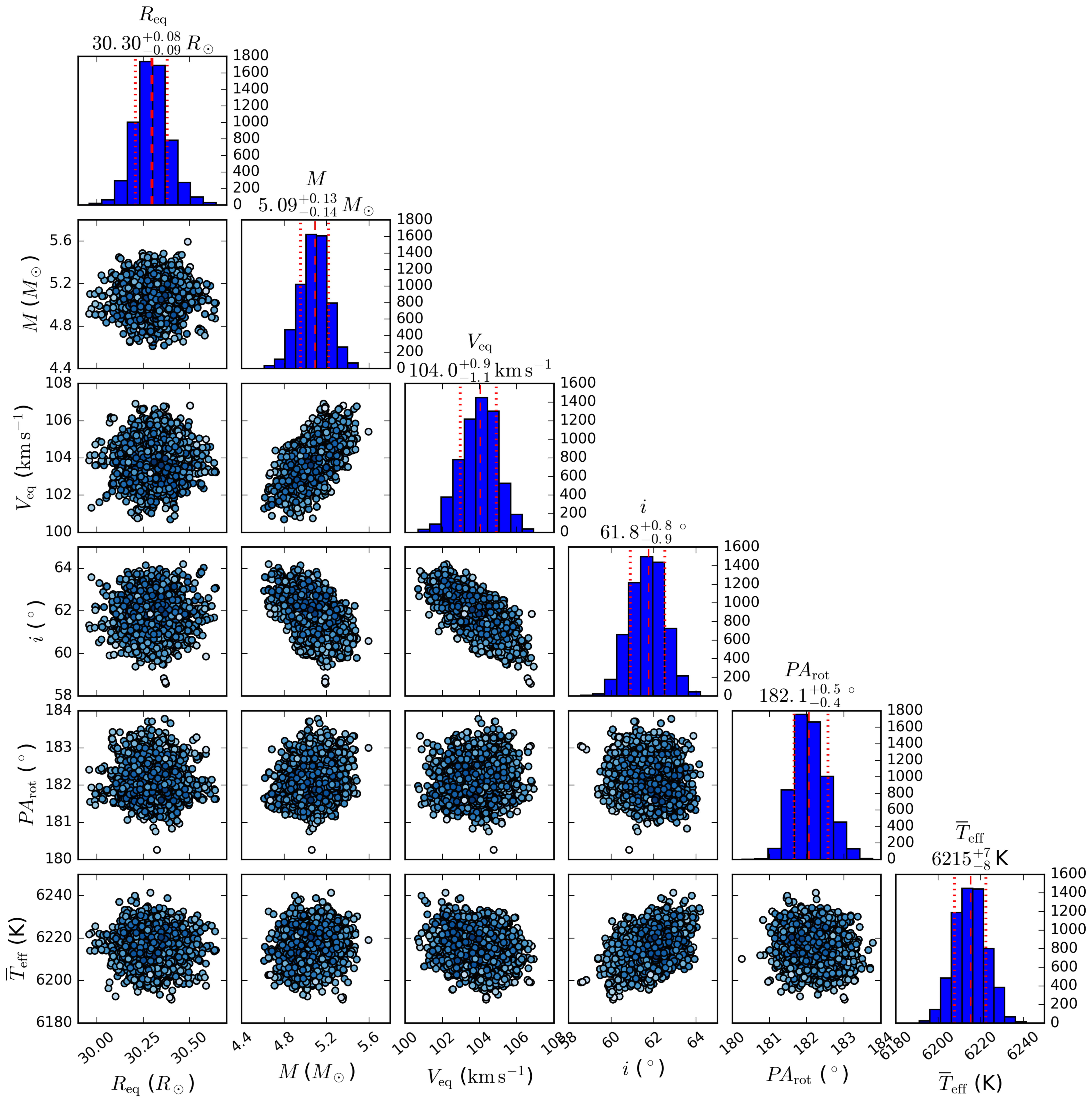}
\caption{As in Fig.~\ref{fig:emcee_histo_results_Rbeta} but for the $\omega$-model.}
\label{fig:emcee_histo_results_RELR}
\end{figure*}

In this section, we constrain the input parameters of the $\beta$- and the $\omega$-models by comparing the CHARRON observables to the real observations of Sargas described in Sect.~\ref{obsdatared}: H-band squared visibilities $V^2$, closure phases $CP$ from PIONIER, and normalized spectral flux $F_\mathrm{norm}$ (around H$\alpha$) from UVES. The use of these three distinct types of observations strengthen the constraints imposed on the model parameters. As discussed in Appendix~\ref{app:band_smear}, the bandwidth smearing effect can be neglected in the interferometric data analyzed in this work.

In a procedure similar to the one adopted by \citet{Domiciano-de-Souza2014_v569pA10}, we performed the comparison model observations through a model-fitting approach using the emcee code \citep{Foreman-Mackey2013_v125p306-312}\footnote{http://dfm.io/emcee/current/}. This code is a Python implementation of the Markov chain Monte Carlo (MCMC) ensemble sampler with affine invariance, which was proposed by \citet{Goodman2010_v5p65-80}.

Given a likelihood function relating model and data, emcee draws samples of the posterior probability density function (PDF), even for large parameter-space dimensions. From these samples one can build histograms of the free parameters of the model and then measure their expectation values and uncertainties. Prior information on the model parameters can also be taken into account by emcee.

In this work, the measurement errors on $V^2$, $CP$, and $F_\mathrm{norm}$ are assumed to be independent and normally distributed, so that the likelihood function is proportional to $\exp(-\chi^2/2)$, where the chi-squared $\chi^2$ has its usual definition.

As described in Sect.~\ref{model}, the input parameters for the $\beta$- and $\omega$-models are $R_\mathrm{eq}$, $M$, $V_\mathrm{eq}$, $\Tmean$, $i$, $\PArot$, $d$, and $\beta$ as an additional parameter for the $\beta$-model. All these quantities were considered as free parameters to be fitted with emcee, except the distance, which was incorporated as a prior, following the value from \citet[][based on revised Hipparcos parallaxes]{Hohle2010_v331p349}: $d = 91.16 \pm 12.51$~pc. Sargas' distance is not given in the Data Release 2 (DR2) of the Gaia mission.


Because $\sim10-20$~s are required for CHARRON to simulate one complete set of Sargas observables on a standard desktop computer, the emcee model-fitting was performed in two steps  so that it could finish in a reasonable time (a few days).

In a first step, emcee runs with a limited number of walkers (100) in a long burn-in and final phases (350 and 50 iterations, resp.). This ensures convergence of emcee within the burn-in phase and allows a first estimate of expectation values and uncertainties $\sigma$. The free parameters were allowed to span a relatively large range of the parameter space. This large, but still limited, parameter space domain was determined based on some preliminary tests with CHARRON and \texttt{LITpro}/JMMC \citep{Tallon-Bosc2008_v7013p70131J}, and on values from previous works. Since not all parameter have been measured in the past or were measured with distinct precisions, they were initialized following uniform distributions over the parameter space domain.

A second step was then performed with a larger number of walkers (400), but shorter burn-in and final phases (100 and 15 iterations resp.). In this step the free parameters were centered on the expected values (median) found in the first step, and were restricted to span a much more limited region of the parameter space (typically from $\sim-7\sigma$ to $\sim7\sigma$ relative to the median), again with initial uniform distributions. Convergence was then still attained during burn-in, and the histograms on the fitted parameters could be built from the final 15 emcee iterations (total of 6000 points). 

The best-fitting parameter values (median) and uncertainties obtained from the CHARRON-emcee fit of the $\beta$- and the $\omega$-models to the observations are presented in Table~\ref{ta:emcee_parameters}. Several derived stellar parameters are also given in this table. We note that Sargas' parameters measured from the $\beta$- and $\omega$-models are in very good agreement (within their uncertainties). 

The direct comparisons between the interferometric and spectroscopic observables and the $\beta$-best-fit model are shown in Figs.~\ref{fig:interf_data_charron_fit_Rbeta} and \ref{fig:flux_data_charron_fit_Rbeta}. The corresponding images for the $\omega$-model are very similar and are given in Appendix~\ref{app:figs_relr} (Figs.~\ref{fig:interf_data_charron_fit_RELR} and \ref{fig:flux_data_charron_fit_RELR}). The signature of rotational flattening is clearly seen in the $V^2$ curves (GD also influences the $V^2$). The $CP$ plots clearly indicate the signature of GD, revealed by the departure from zero seen at high spatial frequencies. 

GD also has a subtle influence on the spectral flux, and in order to better appreciate this we have added the spectra of two $\beta$ models with identical best-fit parameters, but with $T_\mathrm{eff}$ fixed ($\beta$ forced to 0) to the two extreme values $T_\mathrm{p}$ and $T_\mathrm{eq}$. The spectra from these fixed $T_\mathrm{eff}$ models are included in Fig.~\ref{fig:flux_data_charron_fit_Rbeta}, where one can see that they are less effective in reproducing the whole set of spectral features than the complete best-fit model with GD. 

The reduced chi-squared of the best-fit for both models is $\chir=\chi^2/dof=5.4$, which is composed of 0.8 ($V^2$), 3.2 ($CP$), and 1.4 ($F_\mathrm{norm}$), with  2032 ($\beta$-model) and 2033 ($\omega$-model) degrees of freedom ($dof$). The relatively large $\chi^2$ value for the $CP$ comes from somewhat underestimated data error bars, as can be seen in Figs.~\ref{fig:interf_data_charron_fit_Rbeta} or \ref{fig:interf_data_charron_fit_RELR}.

Our MCMC analysis also provides histograms of the fitted parameters and 2D projections of the posterior probability distributions (covariances between parameters). They are shown in Figs.~\ref{fig:emcee_histo_results_Rbeta} and \ref{fig:emcee_histo_results_RELR}, respectively, for the $\beta$- and the $\omega$-models. These 2D projections reveal some correlations between a few parameters. For example a clear correlation appears between $V_\mathrm{eq}$ and $i$, as an expected result of the known difficulty to disentangle these parameters from the strong $V_\mathrm{eq}\sin i$ signature in the data. Correlations are less pronounced in the $\omega$-model. In any case, with or without correlations, the results in Figs.~\ref{fig:emcee_histo_results_Rbeta} and \ref{fig:emcee_histo_results_RELR} show that the model parameters have nicely peaked histograms. 

The results presented in this section show that  both the $\beta$-model and the $\omega$-model can reproduce the observable signatures of fast rotation, with an equivalent description of the surface intensity distribution of Sargas.

\section{Discussion \label{discussion}}



\subsection{Measured parameters}

We discuss below the measured parameters and their consequences for Sargas, also comparing them to previous estimates:

\begin{itemize}
\item  $R_\mathrm{eq}$, $R_\mathrm{p}$, $\overline{R}$, and angular diameters: Our estimated equatorial radius is more precise but compatible with the radius estimation of \citet[][$25.7^{+9.8}_{-7.1}\,\Rsun$]{Samedov1988_v28p335}, considering his uncertainties. His central value is however closer to our polar radius estimation. \citet{Snow1994_v95p163-299} give a much lower value ($17.8\,\Rsun$ without uncertainty), which probably comes from the fact that they used an overly high $T_\mathrm{eff}$ to estimate the radius (see below). The measured equatorial angular diameters ($\diameq$, $\diamp$, and $\overline{\diameter}$) are between the uniform-disk angular-diameter estimates of \citet[][2.2~mas]{Ochsenbein1982_v47p523-531} and \citet[][3.34~mas]{van-Belle2012_v20p51-99}. From the \texttt{GetStar}/JMMC service we obtain $\sim2.64$~mas in the H band, which is similar to our $\diamp$ value.
\item  $M$: Our measured mass is more precise and compatible with the estimates of \citet[][$5.66\pm0.65\,\Msun$]{Hohle2010_v331p349} and \citet[][$6\pm1\,\Msun$]{Samedov1988_v28p335}, considering their uncertainties. Our value is also close to the estimate of \citet[][$5.3\,\Msun$]{Snow1994_v95p163-299}, although they do not provide any uncertainty. This result demonstrates the capability of OLBI to estimate masses of single fast-rotating stars, as also shown by \citet{Zhao2009_v701p209-224} for other targets.
\item $\Tmean$ (and $L$): The measured average $T_\mathrm{eff}$ is lower than the values reported by \citet[][$7268$~K]{Hohle2010_v331p349}, \citet[][$7200$~K]{Snow1994_v95p163-299}, and \citet[][$6750\pm150$~K]{Samedov1988_v28p335}. Their values rather correspond to our polar $T_\mathrm{eff}$, which can be the result of estimation methods more sensitive to the hotter and brighter polar regions. Our derived luminosity $L$ is lower than the estimate of \citet[][$1834\,\Lsun$]{Hohle2010_v331p349}, but closer to those of \citet[][$1230^{+907}_{-522}\,\Lsun$]{Samedov1988_v28p335} and \citet[][$1015\,\Lsun$]{Anderson2012_v38p331-346}. Our measured $\Tmean$ and $L$ suggest a luminosity class around II, according to \citet[][Table~6]{de-Jager1987_v177p217-227}. Considering this luminosity class and Table~5 of \cite{de-Jager1987_v177p217-227}, our estimated $T_\mathrm{p}$, $\Tmean$, and $T_\mathrm{eq}$ correspond, respectively, to spectral types close to \mbox{F2-F4}, \mbox{F5-F6}, and \mbox{F6-F8}. 
\item $V_\mathrm{eq}$ and $i$ (and $\vsini$): We could not find any previous independent measurements of $V_\mathrm{eq}$ and/or $i$ in the literature. Our results are probably the first estimates of these parameters for Sargas. Considering $\vsini$, we found a few spectroscopy-based reported values, but without uncertainties: \citet{Gebocki2005_v560p571} give $125\,\kms$ (tagged with an uncertainty flag), while \citet{Ochsenbein1987_v32p83}, \citet{Hoffleit1991BrightStarCatalogue}, and \citet{Snow1994_v95p163-299} report $105\,\kms$. Our $\vsini$ estimate is close to but lower than these measurements, and is based both on spectroscopy and interferometry (for the first time for this star).
\item $\PArot$: We found no previous measurements of this parameter in the literature for Sargas. However, \citet{Cotton2016_v455p1607-1628} measured a non-negligible degree of polarization on Sargas ($150.8\pm3.3$~ppm), at a position angle of $94.1\degr\pm1.2\degr$, which, considering the uncertainties, is very nearly perpendicular to our $\PArot$. This particular relation between both position angles suggests that the polarization measured on Sargas is intrinsic, supporting the conclusion of \citet{Cotton2016_v455p1607-1628}. In this case, it would be most interesting to perform future theoretical and observational polarimetric studies of Sargas to explore the effects of rotation and GD on this star (and other possible late-type fast rotators). Such studies have been performed by \citet{Cotton2017_v1p690-696}, who obtained compelling results on the B-type fast rotator Regulus. 
\item $\beta$: As far as we know, this work provides the first direct interferometric measurement of GD on Sargas. From Table~\ref{ta:emcee_parameters} one can see that the measured $\beta$ ($\beta$-model) agrees (within $<2\sigma$) with an equivalent $\beta$ derived from the $\omega$-model ($\beta_\omega$). The present work on Sargas therefore provides a new observational validation of the ELR gravity-darkening law. This result constitutes an important verification of their law, since it is the first time that GD is measured with interferometric observations on such a cold, evolved bright-giant, single fast rotator. Considering other GD theories, our measured $\beta$ for Sargas excludes the classical theoretical values of 0.25 \citep{von-Zeipel1924_v84p665-683} and 0.08 \citep{Lucy1967_v65p89-92}, which have been proposed for stars with radiative and convective envelopes, respectively. Considering the more recent theoretical $\beta$ values from \cite{Claret2017_v600pA30} one finds, for stellar parameters close to those of Sargas, that our measured $\beta$ is higher by a factor of $\sim3-4$. Such a discrepancy can come from the fact that Claret's theoretical $\beta$ values appropriate for Sargas assume a convective envelope, which is probably not present in this star (cf. discussion in Sect.~\ref{Sargas_evolution}). Claret's values are also wavelength-dependent and do not correspond to the same wavelengths as in the PIONIER observations.
\end{itemize}

\subsection{Combination of spectroscopy and interferometry \label{Spectro_interfero}}

As far as we know, the present work is the first to simultaneously combine high-resolution spectroscopy and interferometry in a model-fitting analysis of fast rotators. It is therefore interesting to check what information is added by spectroscopy to a good-quality interferometric data set such as the present one. To this end, we performed another MCMC model-fitting, identical to the one described in Sect.~\ref{results}, but including the PIONIER interferometric data alone. Since the model is less constrained because of more limited observational information, a longer final burn-in phase was used in the emcee code (200 iterations) in order to ensure convergence.

The results are shown in Table~\ref{ta:emcee_parameters_no_uves}. As expected, the interferometric data can well constrain the parameters related to size ($R_\mathrm{eq}$) and orientation ($\PArot$), which are in agreement with the analysis using the full data set. On the other hand, the interferometric observables do not strongly impose an absolute scale to the effective temperature (or to the luminosity), resulting in a poorly constrained $\Tmean$ with an overly low value, incompatible with the spectral type of Sargas. This implies that the mass is not constrained either, because GD relates effective temperature and gravity (as e.g., in Eqs.~\ref{eq:teff_rbeta} and \ref{eq:teff_relr}).

It is also interesting to note that although interferometry is sensitive to the apparent flattening, GD can have an opposite influence on the visibilities \citep{Domiciano-de-Souza2002_v393p345-357}, so that both effects need to be measured simultaneously; otherwise there will be a direct impact on the determination of $M$, $\Tmean$, $\beta$, $i$, and $V_\mathrm{eq}$.

The results of this model-fitting experiment show that to correctly measure stellar rotation effects (flattening and GD) from interferometric observations it is important to also ensure a good constraint on the $\teff$ absolute scale. This constraint can be imposed by results from previous works, to be used as prior information, or by photometric data, which was the approach adopted for example by \citet{Zhao2009_v701p209-224}, \citet{Che2011_v732p68-80}, and \citet{Domiciano-de-Souza2014_v569pA10}. In the present work, the constraint on the $\teff$ absolute scale is provided by high-resolution spectroscopy, which, in addition, also provides information on GD and $\vsini$, and finally vastly improves the quality of the solution.


%
%
%
\begin{table}[ht]
\caption[]{\label{ta:emcee_parameters_no_uves} Results from the MCMC model-fitting with emcee, similar to those in Table~\ref{ta:emcee_parameters}, but using the PIONIER interferometric data alone, without the UVES spectroscopic data.}
\centering
\begin{tabular}{ccc}
\hline\hline
\textbf{Fitted parameters}      & \textbf{$\beta$-model}       & \textbf{$\omega$-model} \\
 \multicolumn{3}{c}{\textbf{(Results from interferometric data only)}} \\
\hline
$R_\mathrm{eq}$ $(\Rsun)$   & $30.82^{+0.13}_{-0.12}$       & $31.05^{+0.14}_{-0.15}$     \\
$M$ $(\Msun)$               & $\sim5\pm1.2$                 & $\sim5\pm1.2$ \\
                            & (not constrained)             & (not constrained) \\
$V_\mathrm{eq}$ ($\kms$)    & $103.0^{+12.1}_{-7.7}$        & $105.9^{+12.4}_{-8.0}$   \\
$i$ ($\degr$)                   & $61.9^{+6.0}_{-6.0}$          & $75.6^{+1.0}_{-0.9}$     \\
$\PArot$ ($\degr$)              & $183.1^{+0.5} _{-0.5}$        & $183.3^{+0.5} _{-0.5}$  \\
$\Tmean$ (K)                & $\sim4200\pm250$              & $4536^{+176}_{-149}$ \\ 
                            & (not constrained)             & (poorly constrained) \\
$\beta$                         & $0.147^{+0.016}_{-0.011}$     & $-$   \\
\hline
\end{tabular}
\end{table}

\subsection{Evolutionary status \label{Sargas_evolution}}

\begin{figure*}[ht]
\centering
\includegraphics[width=0.9\hsize]{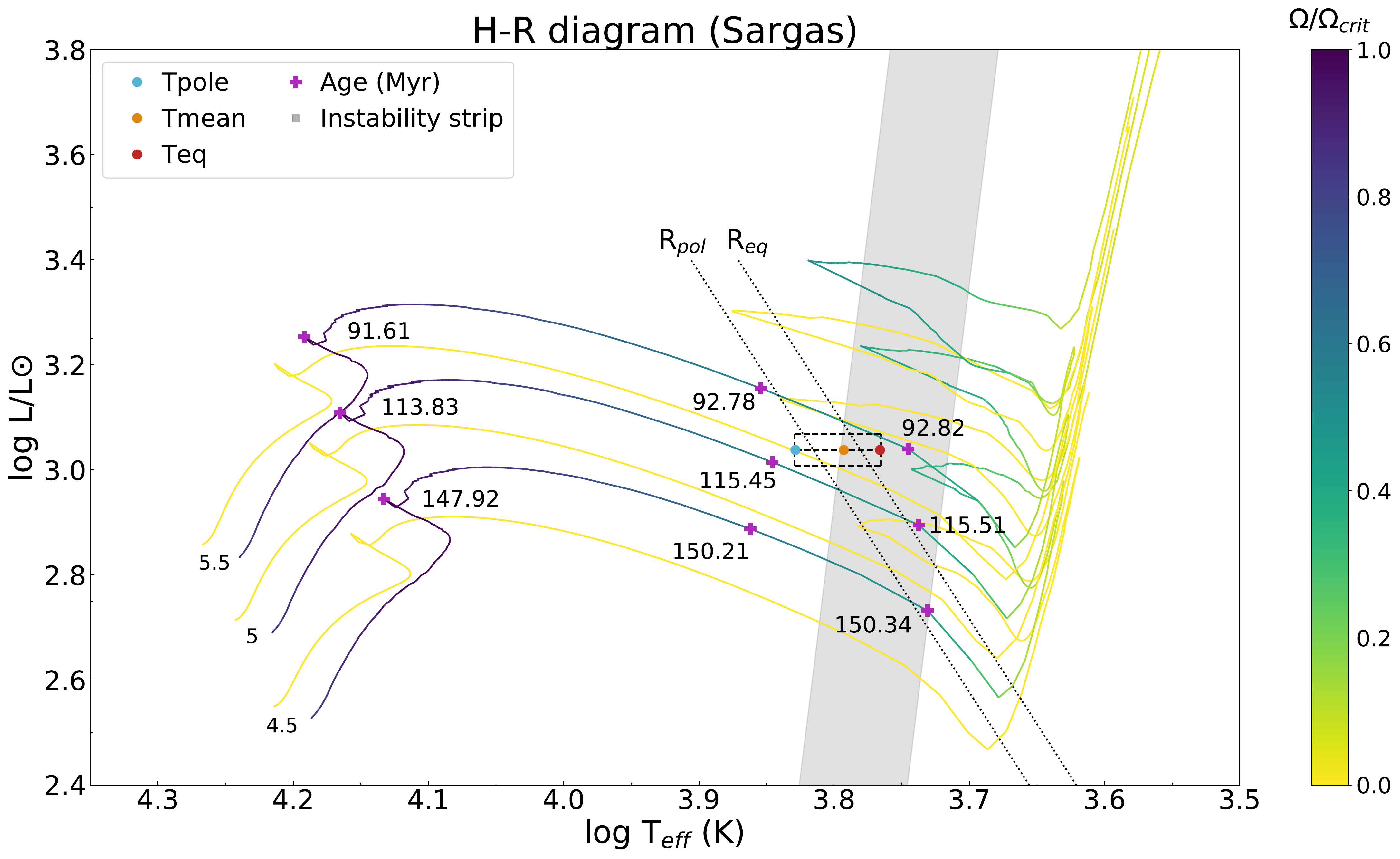}
\caption{Sargas' position in the H-R diagram (best-fit from the $\beta$- and $\omega$-models) together with selected evolutionary tracks from \citet{Georgy2013_v553pA24} at solar metallicity ($\mathrm{Z}=0.014$) for comparison. The selected tracks correspond (1) to initial masses enclosing Sargas' mass, as indicated by the values (in $\Msun$) at the bottom-left (ZAMS position), and (2) to the two extreme rotation rates in Georgy's et al. models, i.e., $\Omega=0.95\Omega_\mathrm{c}$ (multi-colored tracks) and $\Omega=0$ (yellow curves corresponding to evolution without rotation). The position of Sargas is represented by the three colored dots (connected by a dashed line) at the polar ($T_\mathrm{p}$ in blue), average ($\Tmean$ in orange), and equatorial ($T_\mathrm{eq}$ in red) effective temperatures for the estimated luminosity. It is interesting to note that, because of GD, the polar regions of Sargas are located outside the Cepheid instability strip, while the equatorial regions are located inside it; following \citet{Georgy2013_v553pA24} this instability strip was obtained from \citet[][Eq.~37]{Tammann2003_v404p423-448}. We also note that, according to these evolutionary tracks for high rotation rates, Sargas is located in the Hertzsprung gap, corresponding to a short-lived phase of hydrogen-shell burning, just after the end of the main sequence (indicated by crosses and corresponding ages in the left). These high-rotation-rate tracks do not present a blue loop phase compatible with Sargas' position in the H-R diagram. Moreover, the very fast time evolution ($\sim0.1$~Myr) indicated by the ages enclosing (crosses) Sargas' position and the large radius measured by us suggest that this star is quickly approaching the end of the shell-burning phase, just after reaching the Sch{\"o}nberg-Chandrasekhar limit \citep{Schonberg1942_v96p161}. The two dotted straight lines are simply references showing the $L=4\pi R^2\sigma T_\mathrm{eff}^4$ relation for spherical hypothetical stars with radii $R_\mathrm{p}$ and $R_\mathrm{eq}$. This same relation was also used to estimate the uncertainty region for Sargas' position in the H-R diagram (dashed rectangle), where the uncertainties on $L$, $T_\mathrm{p}$, and $T_\mathrm{eq}$ were estimated from the measured uncertainties on $R_\mathrm{eq}$ and $\Tmean$ (cf. Table~\ref{ta:emcee_parameters}).}
\label{fig:hrd_geneva_models_RELR}
\end{figure*}

We now discuss the evolutionary status of Sargas by comparing its stellar parameters determined in this work with theoretical evolutionary tracks. To this end we consider the evolutionary tracks from the Geneva group, computed by \citet{Georgy2013_v553pA24} for rotating stars with solar metallicity. Figure~\ref{fig:hrd_geneva_models_RELR} shows the position of Sargas in the H-R diagram, together with selected evolutionary tracks with similar masses for the highest ($0.95\Omega_\mathrm{c}$) and lowest (no rotation) angular rotation rates on the zero-age main sequence (ZAMS). From this figure, several conclusions can be drawn.

It is presumably certain that Sargas is currently located in the Hertzsprung gap since no track with a high enough rotation rate is compatible with Sargas being in the blue-loop phase. This makes Sargas a very rare star because of the low probability of detecting a star in this short-lived phase ($\sim1-2$~Myr for its mass), in particular considering such a fast rotator. This means that Sargas is presently in the hydrogen-shell-burning phase. Even more remarkable, the very fast evolution time enclosing Sargas' position ($\sim0.1$~Myr as shown in Fig.~\ref{fig:hrd_geneva_models_RELR}) suggests that it is located in the so-called \textit{thin shell burning} phase, that is, it has just reached the Sch{\"o}nberg-Chandrasekhar limit \citep{Schonberg1942_v96p161} and is undergoing a fast core contraction and envelope expansion. During this phase the stellar radius is expected to quickly increase from less than $10\,\Rsun$ to several tens of $\Rsun$, which is totally supported by the large size of Sargas shown by our data. Sargas is probably still fully radiative since it is not yet cool and large enough to have developed a convective envelope, but a more detailed investigation is needed, notably considering the influence of fast-rotation (flattening and GD).

From the measured $\Tmean$ and estimated luminosity it seems that Sargas is located on the hot edge of the Cepheid instability strip. More striking, however, is the fact
that the distinct $T_\mathrm{p}$ and $T_\mathrm{eq}$, resulting from GD, place Sargas' polar regions outside the instability strip, while the equatorial are well inside it, as shown in Fig.~\ref{fig:hrd_geneva_models_RELR} (check also the $T_\mathrm{eff}$ maps in Figs.~\ref{fig:interf_data_charron_fit_Rbeta} and \ref{fig:interf_data_charron_fit_RELR}). Based on this result, it would be interesting in the future to investigate the consequences of this partial location of Sargas on the Cepheid instability strip, notably concerning stellar pulsation.

Determining the age of a rapidly rotating star like Sargas is not a simple task since stellar evolution highly depends on the unknown initial rotation rate, in addition to classical quantities such as initial mass, mass loss, metallicity, overshooting, and also the dimensionality of the model (e.g., 1D, 2D, or 3D). In the framework of the grid of evolutionary tracks adopted here (with a rotation rate at ZAMS of $\Omega_\mathrm{ZAMS}/\Omega_\mathrm{c}=0.95$), and being conservative, we can roughly estimate Sargas' age as $\sim100\pm15$~Myrs. For the adopted evolutionary tracks, this $\Omega_\mathrm{ZAMS}/\Omega_\mathrm{c}$ results in a rotation rate of $\simeq0.5$ at Sargas' age, which is lower than our measured value ($\simeq0.85$). This difference is not surprising since this rotation rate also depends on the quantities mentioned above as well as on the choice of model used, and in turn influences the estimated age. It is interesting to note that using identical models without rotation ($\Omega_\mathrm{ZAMS}=0$ as shown in Fig.~\ref{fig:hrd_geneva_models_RELR}), Sargas' age would be significantly underestimated ($\sim60\%-70\%$ of the estimated value).

\section{Conclusions \label{conclusion}}

In this work we used high-resolution interferometric (VLTI/PIONIER) and spectroscopic (VLT/UVES) data in order to measure several parameters of Sargas, an evolved fast rotator. We used the CHARRON code to compare the $\beta$- and $\omega$-models of rotating stars (described in Sect.~\ref{model}), which adopt the Roche approximation and include the important effects of GD and flattening. From a model-fitting procedure using an MCMC method, we measured several of Sargas' physical parameters, namely $R_\mathrm{eq}$, $M$, $V_\mathrm{eq}$, $i$, $\PArot$, and $\Tmean$, as well as $\beta$, which is required in the $\beta$-model to describe the GD effect.

We showed that both models provide an equivalent description of Sargas, reproducing the observations equally well. This result agrees with those obtained in previous similar works on other fast rotators, and at the same time expands this agreement into a new region of the H-R diagram, never explored in this context of fast-rotation effects: the region of F-type bright giants/supergiants. Indeed, our results validate the $\omega$-model for such stellar types, which is an important step forward in our understanding of stellar rotation effects because, as also discussed by \citet{Espinosa-Lara2011_v533pA43}, this model provides a more physically profound comprehension of GD, without the need for the additional \textit{ad hoc} $\beta$ exponent required in the $\beta$-model.


Thanks to the measured physical parameters of  Sargas we can provide a new, more precise, estimate of its spectral type. We can also guess its evolutionary status by placing it on theoretical evolutionary tracks of the H-R diagram given by the Geneva group \citep{Georgy2013_v553pA24} for rotating stars. This allowed us to estimate Sargas' age and show that it is probably located in the Hertzsprung gap, in the thin-shell-burning phase, just after reaching the Sch{\"o}nberg-Chandrasekhar limit. We have also shown that Sargas is placed very close to the hot edge of the Cepheid instability strip and, because of GD, its equatorial regions are inside this strip, while its polar regions are outside it.

All these intriguing and rare characteristics make Sargas a key target to understand the physics and the evolution of fast rotators of intermediate mass. Although 1D models provide a good starting point, a more profound comprehension of such rapidly rotating stars requires at least a 2D description of their physical structure and evolution. Such a star is obviously a strong motivation to develop 2D stellar models that can follow stellar evolution beyond the main sequence. This is precisely the new challenge of ESTER models \citep[e.g.,][]{Espinosa-Lara2013_v552pA35,Rieutord2016_v318p277-304, gagnier_etal2018}.


\begin{acknowledgements}
This research has made use of the Jean-Marie Mariotti Center (JMMC) services \texttt{OiDB}\footnote{http://oidb.jmmc.fr}, \texttt{OIFits Explorer}\footnote{http://www.jmmc.fr/oifitsexplorer\_page.htm}, \texttt{LITpro}\footnote{http://www.jmmc.fr/litpro}, and  \texttt{GetStar}\footnote{http://www.jmmc.fr/getstar}, co-developed by CRAL, IPAG, and Lagrange/OCA. We are grateful to P.~de~Laverny for providing us the synthetic MARCS spectra from the AMBRE project. We also thanks D.~R.~Reese for his careful reading of the text and suggestions to improve it, and the referee (C.~Haniff) for his critical and constructive remarks. Based on data obtained from the ESO Science Archive Facility under request number 257739. This research made use of the SIMBAD and VIZIER databases (CDS, Strasbourg, France) and NASA's Astrophysics Data System. The authors also acknowledge the strong support of the French Agence Nationale de la Recherche (ANR), under grant ESRR (ANR-16-CE31-0007-01), which made this work possible.
F. Espinosa Lara acknowledges the financial support of the Spanish MINECO
under project ESP2017-88436-R.
\end{acknowledgements}

\bibliographystyle{aa} 
\bibliography{biblio_extracted} 

\begin{appendix}

\section{emcee fit results for the $\omega$-model \label{app:figs_relr}}

\begin{figure*}[ht]  
\centering
 \includegraphics[width=0.46\hsize]{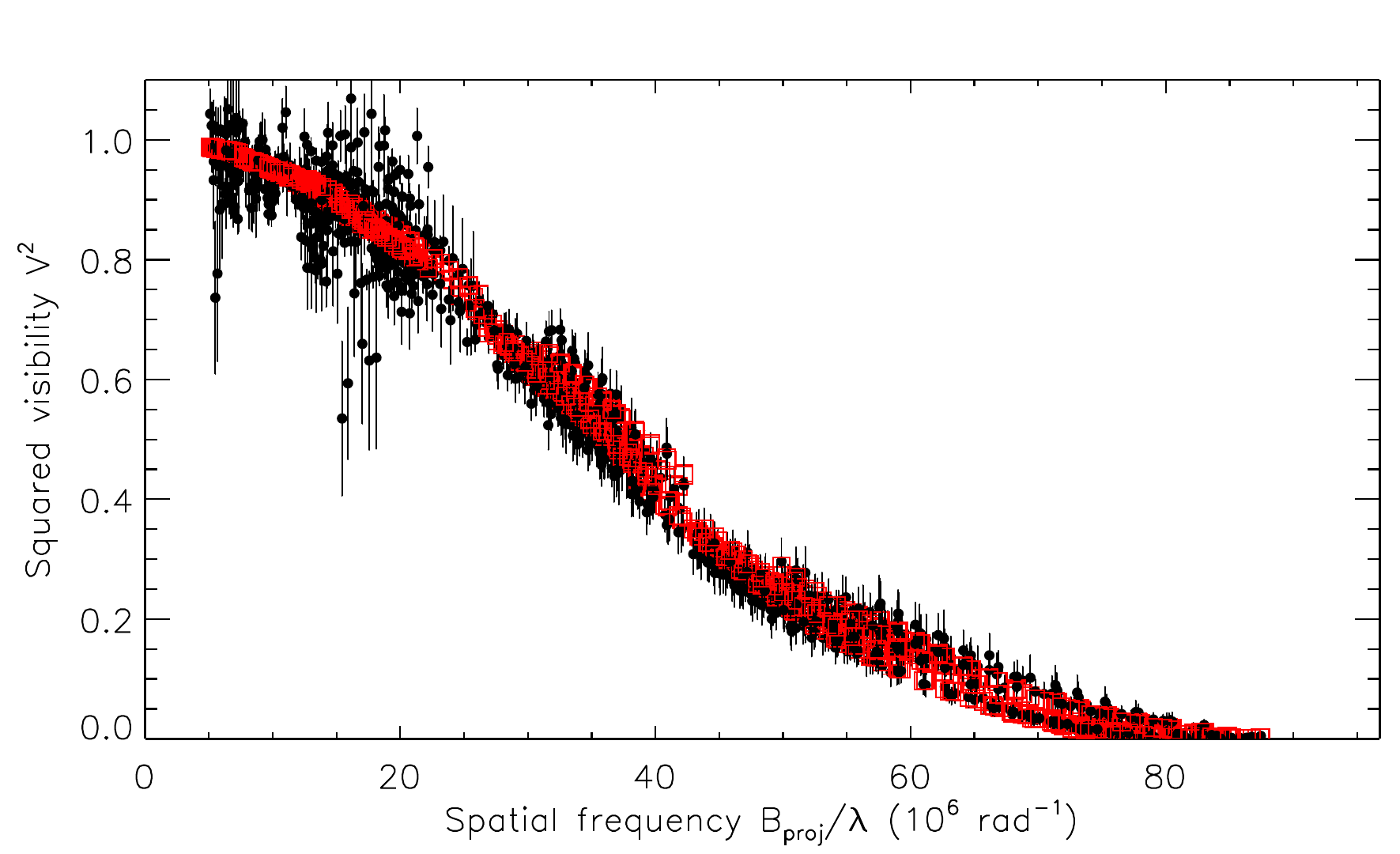}
\hspace*{0.3cm}
  \includegraphics[width=0.46\hsize]{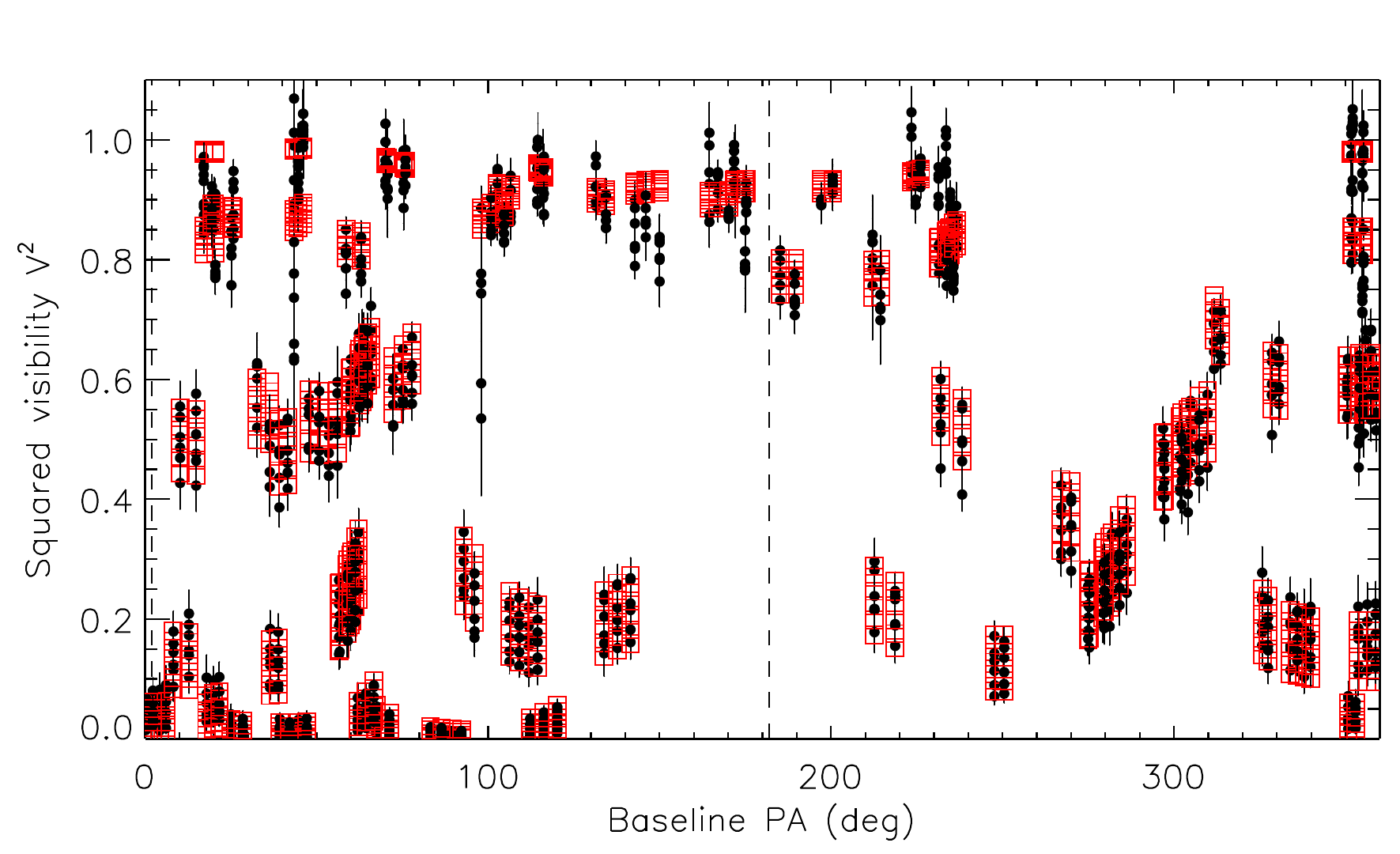}
  \includegraphics[width=0.46\hsize,]{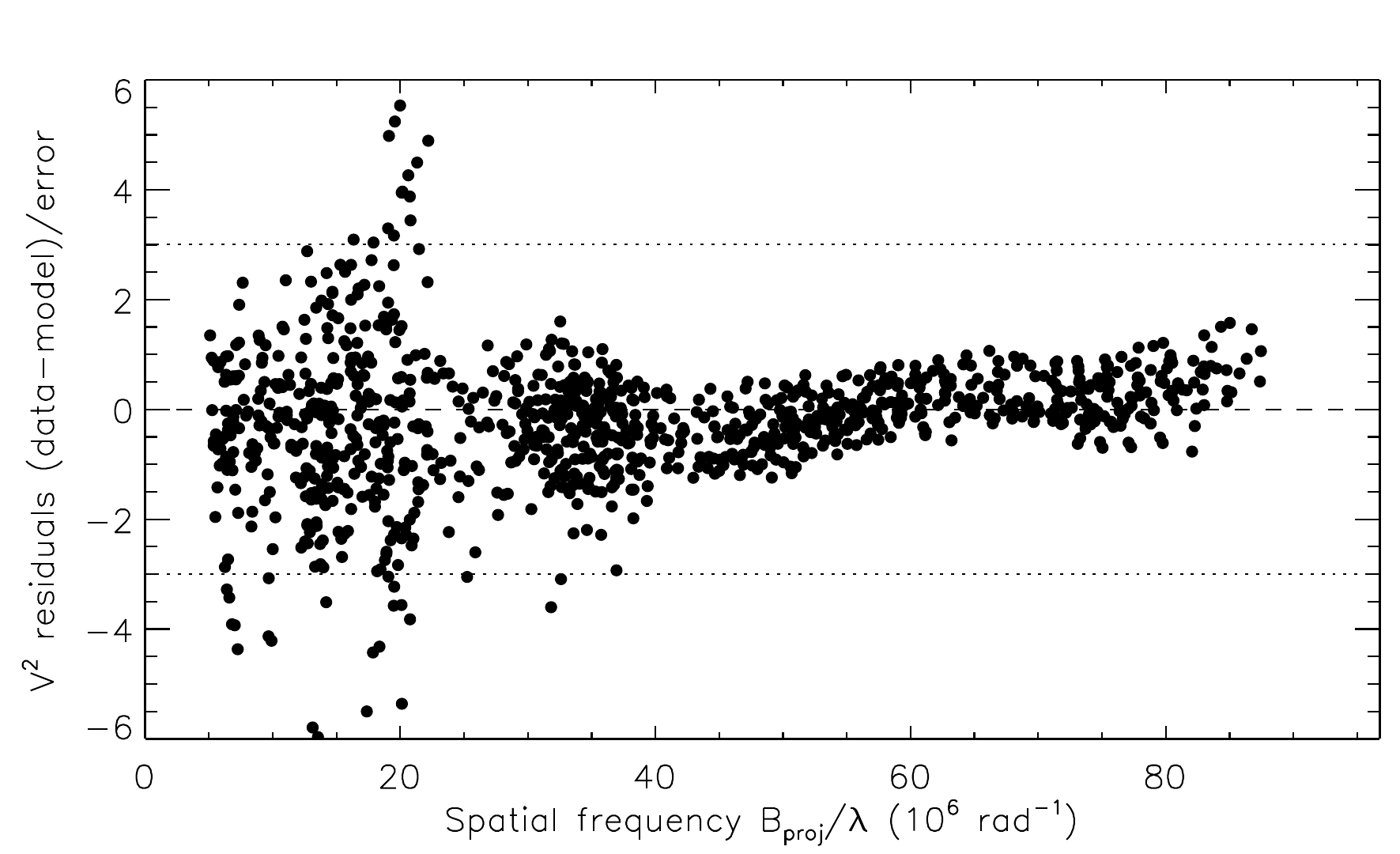}
\hspace*{0.3cm}
  \includegraphics[width=0.46\hsize]{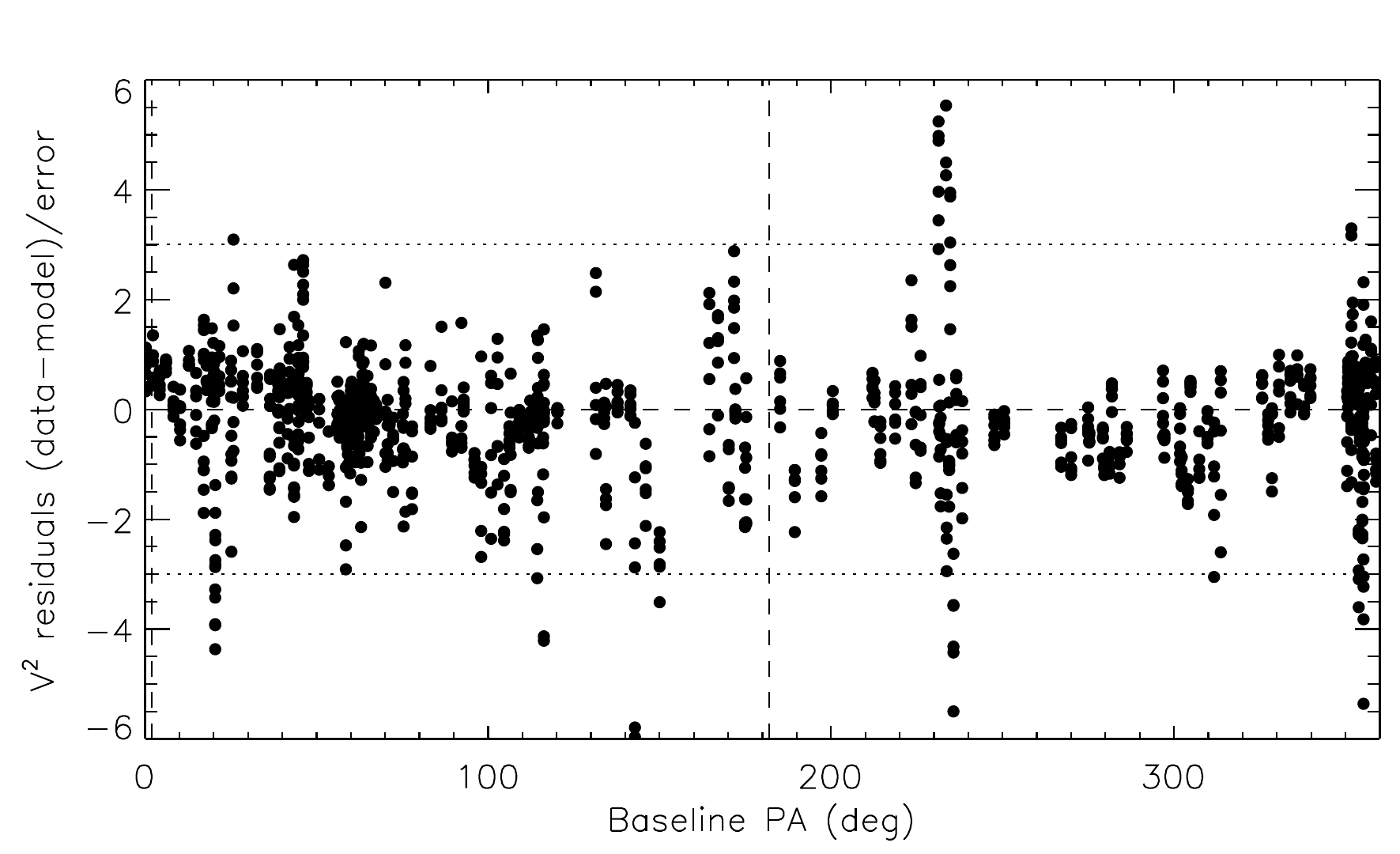}
   \includegraphics[width=0.46\hsize]{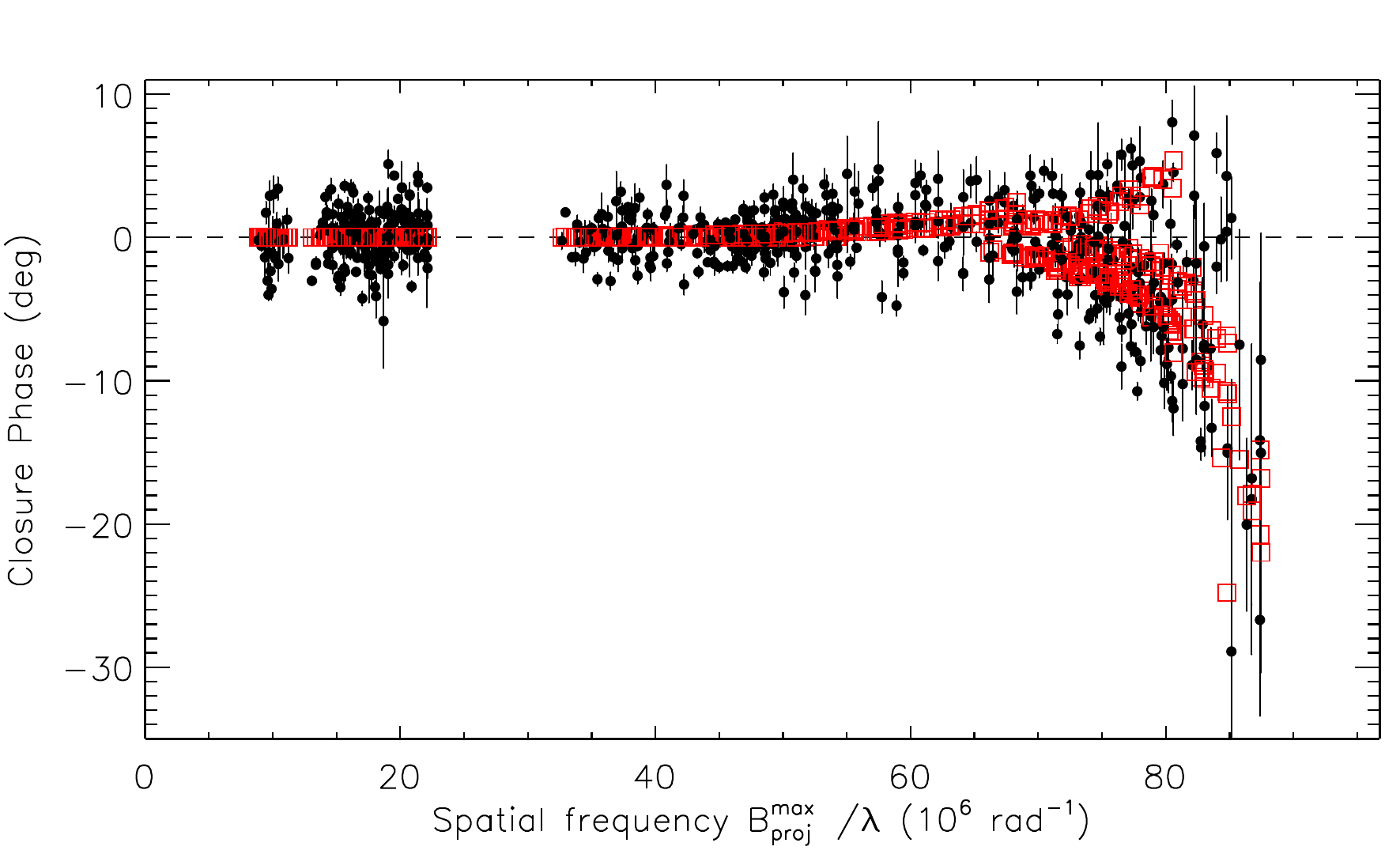}  
\hspace*{1.0cm}
   \includegraphics[width=0.46\hsize]{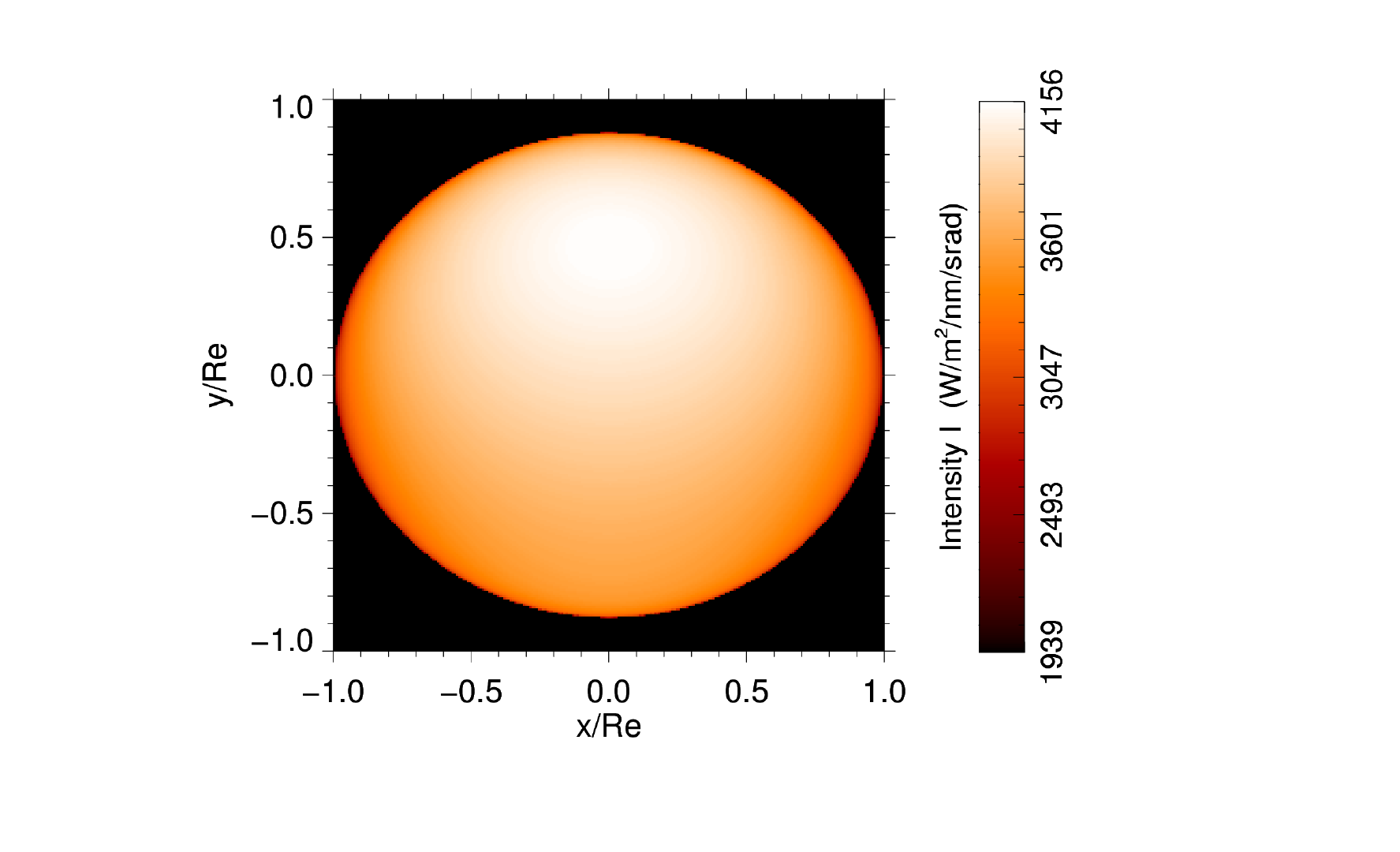}  
   \includegraphics[width=0.46\hsize]{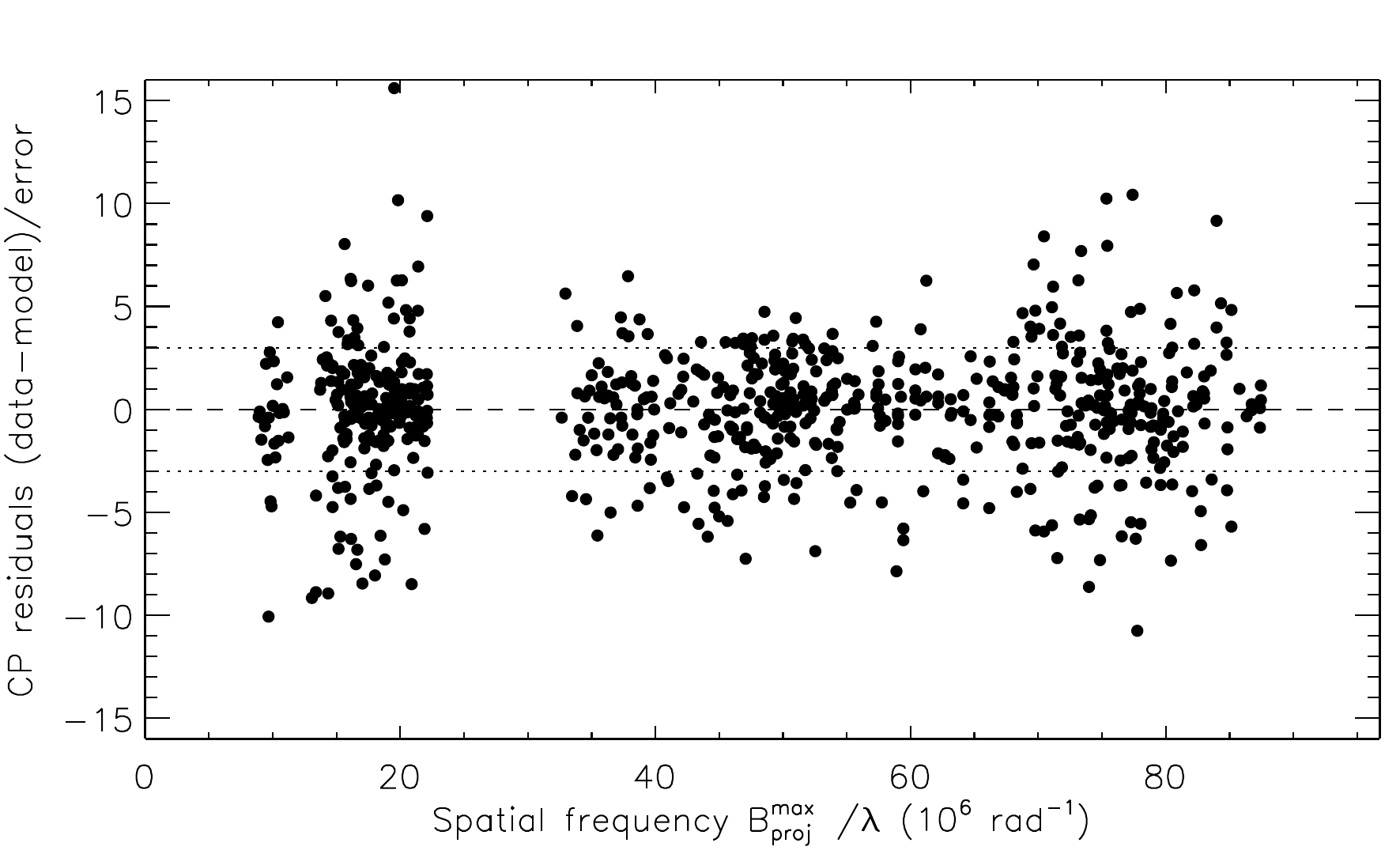}  
\hspace*{1.0cm}
   \includegraphics[width=0.46\hsize]{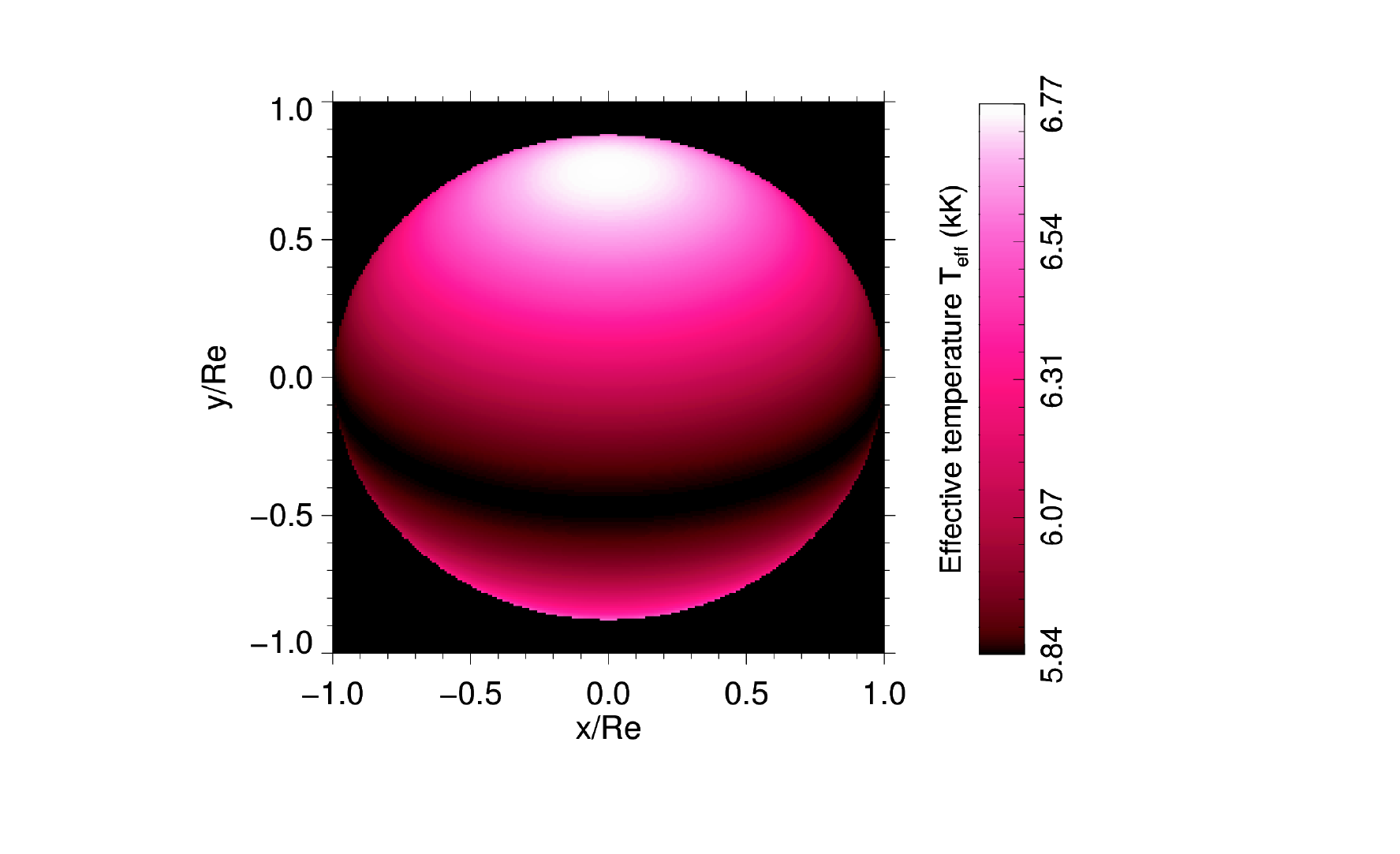}    
\caption{As in Fig.~\ref{fig:interf_data_charron_fit_Rbeta}, but for the comparison of the interferometric observables of PIONIER with the $\omega$-best-fit model (Table~\ref{ta:emcee_parameters}). }
\label{fig:interf_data_charron_fit_RELR}
\end{figure*}
\begin{figure*}[ht]
\centering
\includegraphics[width=0.85\hsize]{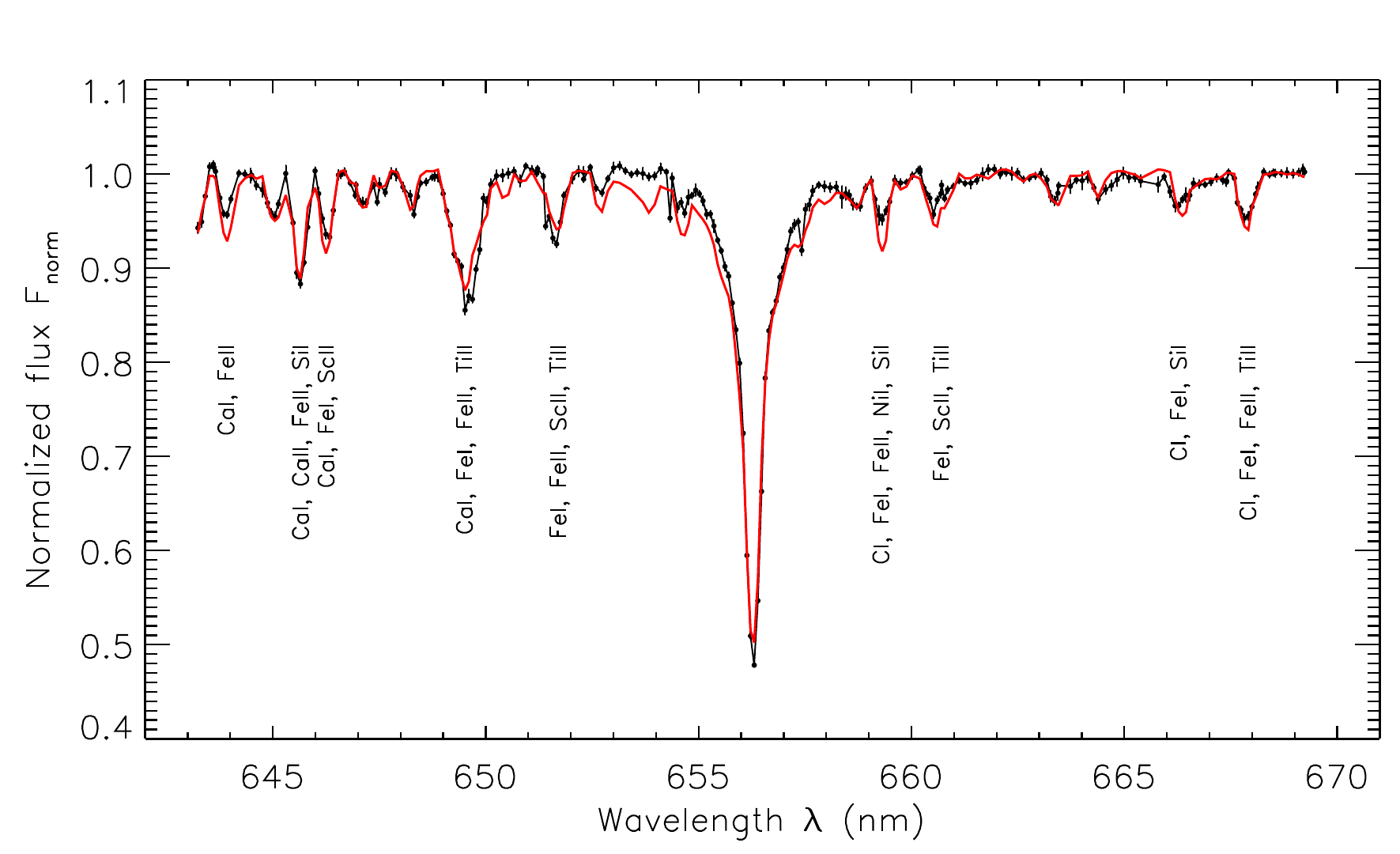}
\caption{As in Fig.~\ref{fig:flux_data_charron_fit_Rbeta}, but for the $\omega$-model. We did not include here the $\beta=0$ (fixed $T_\mathrm{eff}$) spectra since this parameter does not exist in the $\omega$-model. Most of the important spectral lines are well reproduced by our best-fit model.}
\label{fig:flux_data_charron_fit_RELR}
\end{figure*}

\newpage

\section{Bandwidth smearing \label{app:band_smear}}

When analyzing low-spectral-resolution ($R_\mathrm{spec}=\lambda/\Delta\lambda$) interferometric data it is important to check the possible influence of the bandwidth-smearing effect in the corresponding observables. This effect arises from the fact that a finite (broad band) spectral bin contains interferometric information from different spatial frequencies. According to \citet{Davis2000_v318p387-392}, bandwidth smearing is negligible if  
\begin{equation}
 \xi=\frac{\pi \Bproj^\textrm{max} \diameter}{\lambda R_\mathrm{spec}} \ll 1 \ ,
\end{equation}
where $\diameter$ is the typical target's angular size and $\Bproj^\textrm{max}$ is the maximum projected baseline. In the case of PIONIER observations of Sargas considered in this work we have $\xi\sim0.1$, suggesting that the bandwidth smearing effect has a weak influence on the interferometric observables. 

To more firmly confirm this, we computed a model adopting the parameters of our best-fit $\beta$-model (see Sect.~\ref{results}), where the bandwidth smearing effect was included by sub-dividing the PIONIER spectral bins into 25 sub-channels. The interferometric observables computed for each sub-channel were then integrated over the wavelength range of the corresponding PIONIER spectral bin, resulting in observables that include bandwidth smearing and that can be directly compared to the data. This procedure allowed us to confirm that the difference between interferometric observables computed with and without bandwidth smearing are indeed negligible. In particular, the reduced $\chi^2$ are compatible within $\lesssim0.1\%$.

The bandwidth smearing effect can therefore be ignored in the present work, which is a great advantage because computing the interferometric observables on a much larger number of wavelengths would be highly time consuming.


\end{appendix}

\end{document}